\documentclass[twocolumn]{aastex631}

\def\ehtim{\texttt{eht-imaging}}
\def\difmap{\texttt{DIFMAP}}

\usepackage{multirow}
\usepackage[english]{babel}
\usepackage[autostyle, english = american]{csquotes}
\MakeOuterQuote{"}

\defcitealias{Park2021b}{P21}

\begin{document}

\title{Discovery of Limb Brightening in the Parsec-scale Jet of NGC 315 through \\Global Very Long Baseline Interferometry Observations and Its Implications for Jet Models}

\email{jparkastro@khu.ac.kr}

\author[0000-0001-6558-9053]{Jongho Park}
\affiliation{School of Space Research, Kyung Hee University, 1732, Deogyeong-daero, Giheung-gu, Yongin-si, Gyeonggi-do 17104, Republic of Korea}

\author[0000-0002-4417-1659]{Guang-Yao Zhao}
\affiliation{Max-Planck-Institut für Radioastronomie, Auf dem H\"ugel 69, D-53121 Bonn, Germany}

\author[0000-0001-6081-2420]{Masanori Nakamura}
\affiliation{Department of General Science and Education, National Institute of Technology, Hachinohe College, Hachinohe City, Japan}
\affiliation{Institute of Astronomy and Astrophysics, Academia Sinica, P.O. Box 23-141, Taipei 10617, Taiwan, R. O. C.}

\author[0000-0002-8131-6730]{Yosuke Mizuno}
\affiliation{Tsung-Dao Lee Institute, Shanghai Jiao Tong University, 520 Shengrong Road, Shanghai, 201210, China}
\affiliation{School of Physics \& Astronomy, Shanghai Jiao Tong University, 800 Dongchuan Road, Shanghai, 200240, China}

\author[0000-0001-9270-8812]{Hung-Yi Pu}
\affiliation{Department of Physics, National Taiwan Normal University, No. 88, Section 4, Tingzhou Road, Taipei 116, Taiwan, R. O. C.}
\affiliation{Centre of Astronomy and Gravitation, National Taiwan Normal University, No. 88, Section 4, Tingzhou Road, Taipei 116, Taiwan, R. O. C.}

\author[0000-0001-6988-8763]{Keiichi Asada}
\affiliation{Institute of Astronomy and Astrophysics, Academia Sinica, P.O. Box 23-141, Taipei 10617, Taiwan, R. O. C.}

\author[0000-0002-8314-1946]{Kazuya Takahashi}
\affiliation{Research Center for the Early Universe, Graduate School of Science, University of Tokyo, Bunkyo, Tokyo 113-0033, Japan}

\author[0000-0002-7114-6010]{Kenji Toma}
\affiliation{Frontier Research Institute for Interdisciplinary Sciences, Tohoku University, Sendai 980-8578, Japan}
\affiliation{Astronomical Institute, Graduate School of Science, Tohoku University, Sendai 980-8578, Japan}

\author[0000-0002-2709-7338]{Motoki Kino}
\affiliation{Kogakuin University of Technology \& Engineering, Academic Support Center, 2665-1 Nakano-machi, Hachioji, Tokyo 192-0015, Japan}
\affiliation{National Astronomical Observatory of Japan, Osawa 2-21-1, Mitaka, Tokyo 181-8588, Japan}

\author[0000-0001-6083-7521]{Ilje Cho}
\affiliation{Korea Astronomy and Space Science Institute, Daedeok-daero 776, Yuseong-gu, Daejeon 34055, Republic of Korea}
\affiliation{Yonsei University, Department of Astronomy, Seoul, Republic of Korea}
\affiliation{Instituto de Astrof\'{i}sica de Andaluc\'{i}a—CSIC, Glorieta de la Astronom\'{i}a s/n, E-18008 Granada, Spain}

\author[0000-0001-6906-772X]{Kazuhiro Hada}
\affiliation{Graduate School of Science, Nagoya City University, Yamanohata 1, Mizuho-cho, Mizuho-ku, Nagoya, 467-8501, Aichi, Japan}
\affiliation{Mizusawa VLBI Observatory, National Astronomical Observatory of Japan, 2-12 Hoshigaoka-cho, Mizusawa, Oshu, 023-0861, Iwate, Japan.}

\author[0000-0002-8186-4753]{Phil G. Edwards}
\affiliation{Australia Telescope National Facility, CSIRO Astronomy and Space Science, P.O. Box 76, Epping NSW 1710, Australia}

\author[0000-0002-7322-6436]{Hyunwook Ro}
\affiliation{Korea Astronomy and Space Science Institute, Daedeok-daero 776, Yuseong-gu, Daejeon 34055, Republic of Korea}

\author[0000-0001-9799-765X]{Minchul Kam}
\affiliation{Department of Physics and Astronomy, Seoul National University, Gwanak-gu, Seoul 08826, Republic of Korea}

\author[0009-0007-8554-4507]{Kunwoo Yi}
\affiliation{Department of Physics and Astronomy, Seoul National University, Gwanak-gu, Seoul 08826, Republic of Korea}

\author{Yunjeong Lee}
\affiliation{Department of Astronomy and Space Science, Kyung Hee University, 1732, Deogyeong-daero, Giheung-gu, Yongin-si, Gyeonggi-do 17104, Republic of Korea}

\author[0000-0002-3723-3372]{Shoko Koyama}
\affiliation{Graduate School of Science and Technology, Niigata University,
8050 Ikarashi 2-no-cho, Nishi-ku, Niigata 950-2181, Japan}
\affiliation{Institute of Astronomy and Astrophysics, Academia Sinica, P.O. Box 23-141, Taipei 10617, Taiwan, R. O. C.}

\author[0000-0003-1157-4109]{Do-Young Byun}
\affiliation{Korea Astronomy and Space Science Institute, Daedeok-daero 776, Yuseong-gu, Daejeon 34055, Republic of Korea}
\affiliation{University of Science and Technology, Gajeong-ro 217, Yuseong-gu, Daejeon 34113, Republic of Korea}

\author[0000-0002-5851-5264]{Chris Phillips}
\affiliation{Australia Telescope National Facility, CSIRO Astronomy and Space Science, P.O. Box 76, Epping NSW 1710, Australia}

\author[0000-0002-8978-0626]{Cormac Reynolds}
\affiliation{Australia Telescope National Facility, CSIRO Astronomy and Space Science, P.O. Box 76, Epping NSW 1710, Australia}

\author[0000-0001-6094-9291]{Jeffrey A. Hodgson}
\affiliation{Department of Physics and Astronomy, Sejong University, 209 Neungdong-ro, Gwangjin-gu, Seoul 05006, Republic of Korea}

\author[0000-0002-6269-594X]{Sang-Sung Lee}
\affiliation{Korea Astronomy and Space Science Institute, Daedeok-daero 776, Yuseong-gu, Daejeon 34055, Republic of Korea}
\affiliation{University of Science and Technology, Gajeong-ro 217, Yuseong-gu, Daejeon 34113, Republic of Korea}

\begin{abstract}

We report the first observation of the nearby giant radio galaxy NGC 315 using a global VLBI array consisting of 22 radio antennas located across five continents, including high-sensitivity stations, at 22 GHz. Utilizing the extensive $(u,v)$-coverage provided by the array, coupled with the application of a recently developed super-resolution imaging technique based on the regularized maximum likelihood method, we were able to transversely resolve the NGC 315 jet at parsec scales for the first time. Previously known for its central ridge-brightened morphology at similar scales in former VLBI studies, the jet now clearly exhibits a limb-brightened structure. This finding suggests an inherent limb-brightening that was not observable before due to limited angular resolution. Considering that the jet is viewed at an angle of $\sim50^\circ$, the observed limb-brightening is challenging to reconcile with the magnetohydrodynamic models and simulations, which predict that the Doppler-boosted jet edges should dominate over the non-boosted central layer. The conventional jet model that proposes a fast spine and a slow sheath with uniform transverse emissivity may pertain to our observations. However, in this model, the relativistic spine would need to travel at speeds of $\Gamma\gtrsim6.0-12.9$ along the de-projected jet distance of (2.3--10.8)$\times 10^3$ gravitational radii from the black hole. We propose an alternative scenario that suggests higher emissivity at the jet boundary layer, resulting from more efficient particle acceleration or mass loading onto the jet edges, and consider prospects for future observations with even higher angular resolution.

\end{abstract}


\section{Introduction}
\label{sec:introduction}

Active Galactic Nuclei (AGNs) often act as hosts to tightly collimated relativistic jets, with the typical parsec-scale morphology characterized by a compact, optically thick core and an extended jet \citep[e.g.,][]{Lister2018, Weaver2022}. The extended jet typically exhibits a central ridge-brightened morphology, meaning that the transverse intensity is greatest along the jet axis. However, certain AGN jets exhibit limb-brightening phenomena, observed notably in the parsec-scale jets of nearby radio galaxies and blazars\footnote{It should be noted that a recent study has claimed limb-brightening in the nearby quasar 1928+738 \citep{Yi2024}.} such as M87 \citep{Walker2018, Lu2023}, Cygnus A \citep{Boccardi2016a, Boccardi2016b}, 3C 84 \citep{Nagai2014, Giovannini2018, Savolainen2023}, 3C 264 \citep{Boccardi2019}, 3C 273 \citep{Bruni2021}, as well as Mrk 421 \citep{Piner2010} and Mrk 501 \citep{Giroletti2004, Piner2009, Koyama2019}.

A recent observation of the nearby radio galaxy Centaurus A, conducted with the Event Horizon Telescope (EHT; \citealt{EHT2019a, EHT2022a, EHT2024}) at 1.3\,mm, has unveiled a limb-brightened jet morphology \citep{Janssen2021}. However, at parsec scales, previous VLBI observations at centimeter wavelengths have indicated a central ridge-brightened morphology for the jet \citep{Horiuchi2006, Ojha2010, Muller2014}. A plausible explanation for this difference is that the jet is inherently limb-brightened, but this characteristic has not been discernible due to the limited angular resolution of previous centimeter VLBI observations.

In light of this finding, an intriguing question emerges: Could all AGN jets possess an intrinsic limb-brightened nature on parsec scales? Notably, the jets reported to exhibit limb-brightening characteristics at this scale are associated with nearby AGNs and/or substantial black hole masses. For instance, consider the luminosity distance ($d_L$) and black hole mass ($M_{\rm BH}$) of the sources showcasing limb-brightened jets: M87 ($d_L = 16.8$ Mpc; $M_{\rm BH} = 6.5\times10^9 M_\odot$; \citealt{Gebhardt2011, EHT2019f}; 1 milliarcsecond $\sim2.6\times10^2\ R_g$, where $R_g$ is the gravitional radius), Cygnus A ($d_L = 232$ Mpc; $M_{\rm BH} = 2.5\times10^9 M_\odot$; \citealt{Tadhunter2003}; 1 mas $\sim8.4\times10^3\ R_g$), 3C 84 ($d_L = 76.9$ Mpc; $M_{\rm BH} = 8\times10^8 M_\odot$; \citealt{Scharwachter2013}; 1 mas $\sim9.4\times10^3\ R_g$), 3C 264 ($d_L = 94$ Mpc; $M_{\rm BH} = 4.7\times10^8 M_\odot$; \citealt{Balmaverde2008}; 1 mas $\sim1.9\times10^4\ R_g$), Mrk 421 ($d_L = 127$ Mpc; $M_{\rm BH} = 9.0\times10^8 M_\odot$; \citealt{Lico2017}; 1 mas $\sim1.3\times10^4\ R_g$), Mrk 501 ($d_L = 147$ Mpc; $M_{\rm BH} = (0.9-3.4)\times10^9 M_\odot$; \citealt{Barth2002}; 1 mas $\sim(4.1-15.5)\times10^3\ R_g$), and Centaurus A ($d_L = 4$ Mpc; $M_{\rm BH} = 5.5\times10^7 M_\odot$; \citealt{Neumayer2010}; 1 mas $\sim7.4\times10^3\ R_g$). These jets are readily resolvable in their transverse structures\footnote{Typically, the jet width scales with the black hole mass; e.g., \citealt{Kovalev2020, Boccardi2021, Park2021b}, and it is easier to resolve the jets transversely for sources having larger black hole masses.}. A plausible explanation for this pattern could be as follows: while many, if not all, AGN jets potentially exhibit limb-brightening on parsec scales, the majority of these jets have been known to manifest central ridge-brightening characteristics due to limitations in angular resolution in previous VLBI observations.

NGC 315 (J0057+3021), a giant elliptical galaxy in close proximity, is situated at a redshift of 0.01648 \citep{Trager2000}. It is classified as a low-luminosity AGN (LLAGN; \citealt{Ho2008}) with the mass accretion rate inferred to be $\approx10^{-4}\dot{M}_{\rm Edd}$, where $\dot{M}_{\rm Edd}$ is the Eddington mass accretion rate \citep{Gu2007, Worrall2007, Morganti2009, Ricci2022}. It harbors a radio source of Fanaroff-Riley type I (FR I; \citealt{FR1974}), featuring two-sided jets that extend over several hundred kiloparsecs from the core of the galaxy \citep[e.g.,][]{Bridle1976, Laing2006}. Detailed examinations of these kpc-scale jets have been carried out using observations conducted with the Very Large Array (VLA). The primary findings can be briefly outlined as follows: (i) the jet viewing angle is $\approx50^\circ$ \citep{LB2014}, (ii) significant deceleration is observed in the jets \citep{Canvin2005, LB2014}, (iii) the velocity field of the jets exhibits a stratified structure laterally, with the outer regions displaying slower speeds compared to the on-axis regions, and (iv) the radio spectral index is flatter at the jet's edge than along its axis within the regions of deceleration and downstream.

Our recent work involved multifrequency Very Long Baseline Array (VLBA) observations, coupled with a comprehensive analysis of archival High Sensitivity Array (HSA) and VLA data (\citealt{Park2021b}; hereafter \citetalias{Park2021b}). We discovered that the jet's geometry undergoes a transition from a semi-parabolic configuration to a conical/hyperbolic shape at a distance of $\sim 10^5\ R_g$. Additionally, our findings indicate that the parsec-scale jets experience a gradual acceleration, reaching relativistic speeds within the zone where jet collimation takes place. This supports the presence of the jet acceleration and collimation zone predicted in previous theoretical studies \citep[e.g.,][]{VK2004, Marscher2008, Meier2012}. These behaviors closely resemble those observed in the M87 jet \citep[e.g.,][]{AN2012, Hada2013, NA2013, Asada2014, Mertens2016, Nakamura2018, Park2019b}. We note that other recent investigations utilizing the VLBA/HSA and VLA have proposed that jet collimation and acceleration occur at parsec scales, but over distances about an order of magnitude smaller than those reported in our study \citep{Boccardi2021, Ricci2022}.

Earlier VLBI observations have indicated that the jets of NGC 315 exhibit a central ridge-brightened morphology on parsec scales \citep[e.g.,][]{Venturi1993, Cotton1999, Boccardi2021, Ricci2022}. In our recent study \citepalias{Park2021b}, we observed a similar phenomenon, albeit with an interesting deviation in the image captured using the HSA at 43 GHz. This image reveals a limb-brightened morphology within a limited jet distance range. It is important to note that the image was convolved with a circular beam sized equivalent to the minor axis of the synthesized beam, and the limb-brightening was indicated only in a super-resolution image. Therefore, the robustness of the limb-brightening observed in the jet remained uncertain.

In this paper, we utilize an observation conducted through a global VLBI array to test our hypothesis regarding the nature of NGC 315 jets on parsec scales. The array consists of stations from four local VLBI networks: the European VLBI Network (EVN\footnote{\url{https://evlbi.org/}}), the VLBA\footnote{\url{https://science.nrao.edu/facilities/vlba}}, the East Asia VLBI Network (EAVN\footnote{\url{https://radio.kasi.re.kr/eavn/main.php}}; \citealt{Cui2021, Cui2023, EAVN2022, Cho2022}), and the Australian Long Baseline Array (LBA\footnote{\url{https://www.atnf.csiro.au/vlbi/overview/index.html}}; \citealt{EP2015}).

The primary motivation for using the global array is to robustly resolve the jet transversely by enhancing the angular resolution and sensitivity of the data, thus drawing a firm conclusion on the limb-brightening hinted at by previous observations. In this paper, we will refer to the array as the Global VLBI Array (GVA), as there is currently no official name for it. The efforts to combine various VLBI arrays have recently been maintained by the Global VLBI Alliance\footnote{\url{http://gvlbi.evlbi.org}} \citep{VLBI2030}, facilitating the flow of information between VLBI networks, including sharing strategies, technical developments for compatibility, logistics, operations, and user support.

While there have been past observations utilizing the GVA \citep[e.g.,][]{Bruni2017, Bruni2021, Vegagarcia2020, Johnson2021, Baczko2022, Kim2023}, many of these were part of space-VLBI endeavors for the RadioAstron mission, and the full potential of the GVA has yet to be fully unveiled and explored. One notable example illustrating the potential of the GVA is presented in our recent study \citep{Park2024}, where we demonstrated an indication of a dense ambient medium shaping the jet in the parsec-scale jet of Perseus A, thanks to the very high angular resolution achieved by the GVA.

The paper is structured as follows: In Section~\ref{sec:observations}, we provide details about the observations and the data reduction process. Section~\ref{sec:analysis} presents the results of our analysis regarding limb brightening of the jet as well as the jet's collimation profile. This is followed by Section~\ref{sec:discussion}, where we discuss the obtained results. We conclude with future prospects in Section~\ref{sec:conclusion}. For our calculations, we adopt an angular diameter distance of 71.832 Mpc for NGC 315, derived from the following cosmological parameters: $H_0 = 67.4$, $\Omega_m = 0.315$, and $\Omega_{\Lambda} = 0.685$ \citep{Planck2020}.

\section{Observations and Data Reduction}
\label{sec:observations}

In this section, we provide a concise overview of the GVA observation, data reduction, and imaging procedures. For a more comprehensive explanation of these procedures, see Appendix~\ref{appendix:observations}.

\begin{figure}[t]
\centering
\includegraphics[width=\linewidth]{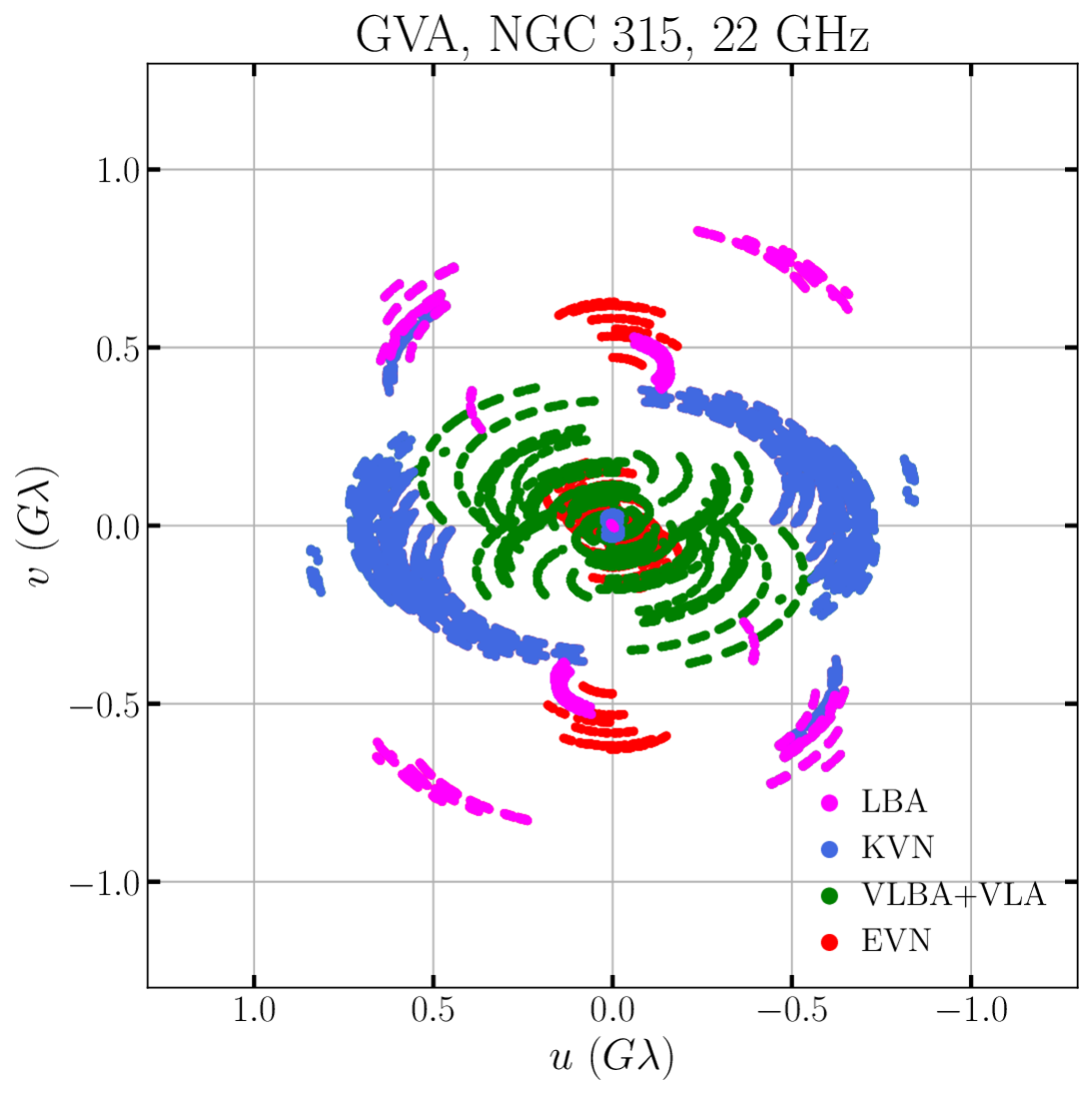}
\caption{The $(u,v)$-coverage for the GVA 22 GHz data is depicted, emphasizing baselines from LBA, KVN, VLBA+VLA, and EVN in magenta, blue, green, and red colors, respectively. The data have been averaged across the entire bandwidth.}
\label{fig:uvcoverage}
\end{figure}

On November 9, 2022, we observed NGC 315 at 22 GHz using the GVA. This observation involved 22 radio telescopes spread across five continents. We applied standard data reduction techniques with the NRAO's Astronomical Image Processing System (AIPS; \citealt{Greisen2003}). The $(u,v)$-coverage of the fringe-detected visibilities is presented in Figure~\ref{fig:uvcoverage}. For imaging the data, we employed two approaches: inverse modeling with CLEAN \citep{Hogbom1974, Clark1980, RC2011}, using the \difmap{} software package \citep{Shepherd1997}, and forward modeling with the regularized maximum likelihood (RML) method implemented in the \ehtim{} library\footnote{\url{https://github.com/achael/eht-imaging}} \citep{Chael2016, Chael2018, Chael2023}. We iteratively performed CLEAN and self-calibration according to standard procedures.

For RML imaging with \ehtim{}, we adopted a strategy similar to that used in the \emph{RadioAstron} space-based VLBI observations of Perseus A \citep{Savolainen2023}. The final images are influenced by the selection of regularizers and their weights. Consequently, we conducted a small parameter survey for four regularizer terms: the maximum entropy method (MEM), total variation (TV), total squared variation (TV2), and the $l1$ norm. We compared 10 images that provided the lowest reduced $\chi^2$ for the closure quantities and concluded that \ehtim{} can achieve an effective spatial resolution of 30\% of the diffraction limit (see Appendix~\ref{appendix:survey} for more details). Therefore, all RML images presented in this study were blurred with the specified blurring kernel. We selected the image with the overall minimum reduced $\chi^2$ for the closure quantities as the representative image.

\begin{figure*}[t]
\centering
\includegraphics[width=\linewidth]{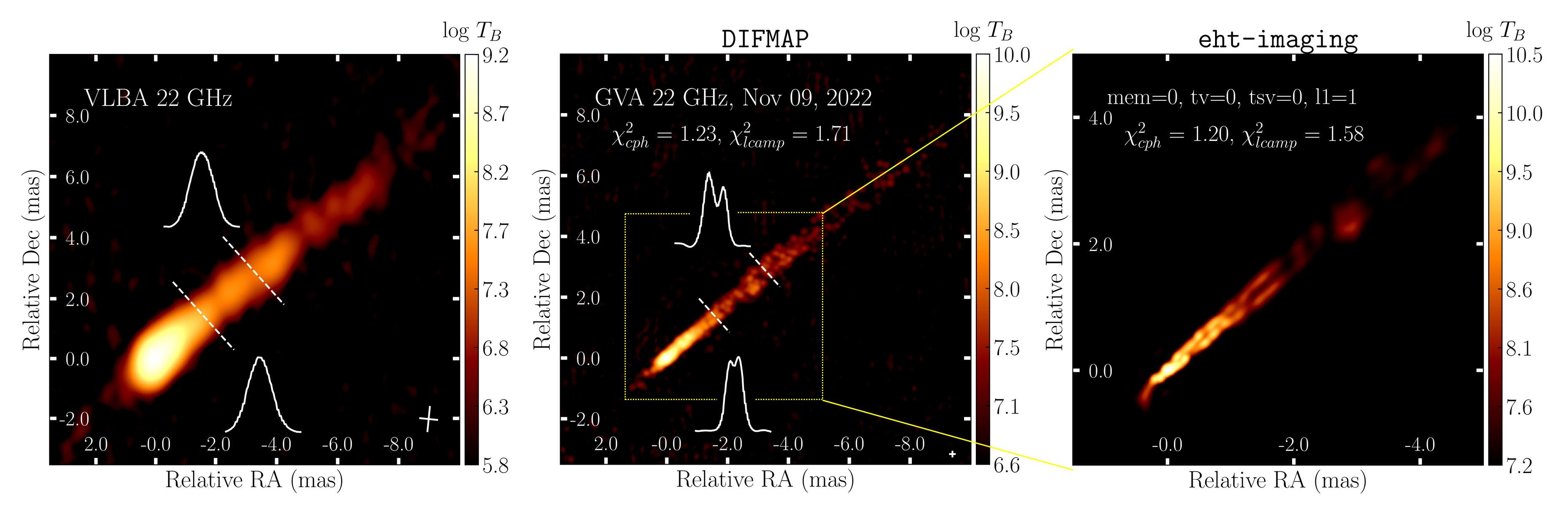}
\caption{Images of the NGC 315 jets at parsec and subparsec scales. \emph{Left:} CLEAN image obtained with \difmap{} using VLBA data at 22 GHz on January 5, 2020, as published in \citetalias{Park2021b}. The restoring beam size is indicated in the bottom right corner. Intensity profiles along a transverse direction to the jet axis at two locations (along the white dashed lines) are presented. \emph{Middle:} CLEAN image obtained with \difmap{} using GVA data at 22 GHz on November 9, 2022. The CLEAN model is restored with a circular Gaussian beam, with a size of the minor axis FWHM of the synthesized beam, approximately $\sim0.17$ mas. A limb-brightened jet morphology is observed at various locations, not visible in the lower-resolution VLBA-only image on the left. \emph{Right:} Representative RML image obtained with \ehtim{} using the GVA data. The field-of-view corresponds to the yellow dotted box in the middle figure. The image was blurred using a circular Gaussian blurring kernel with a FWHM that is 30\% of the diffraction limit (see Appendix~\ref{appendix:survey}). The jet structure is clearly resolved transversely, revealing a limb-brightened jet morphology at sub-mas and mas scales.}
\label{fig:fiducial}
\end{figure*}

\begin{figure}[t]
\centering
\includegraphics[width=\linewidth]{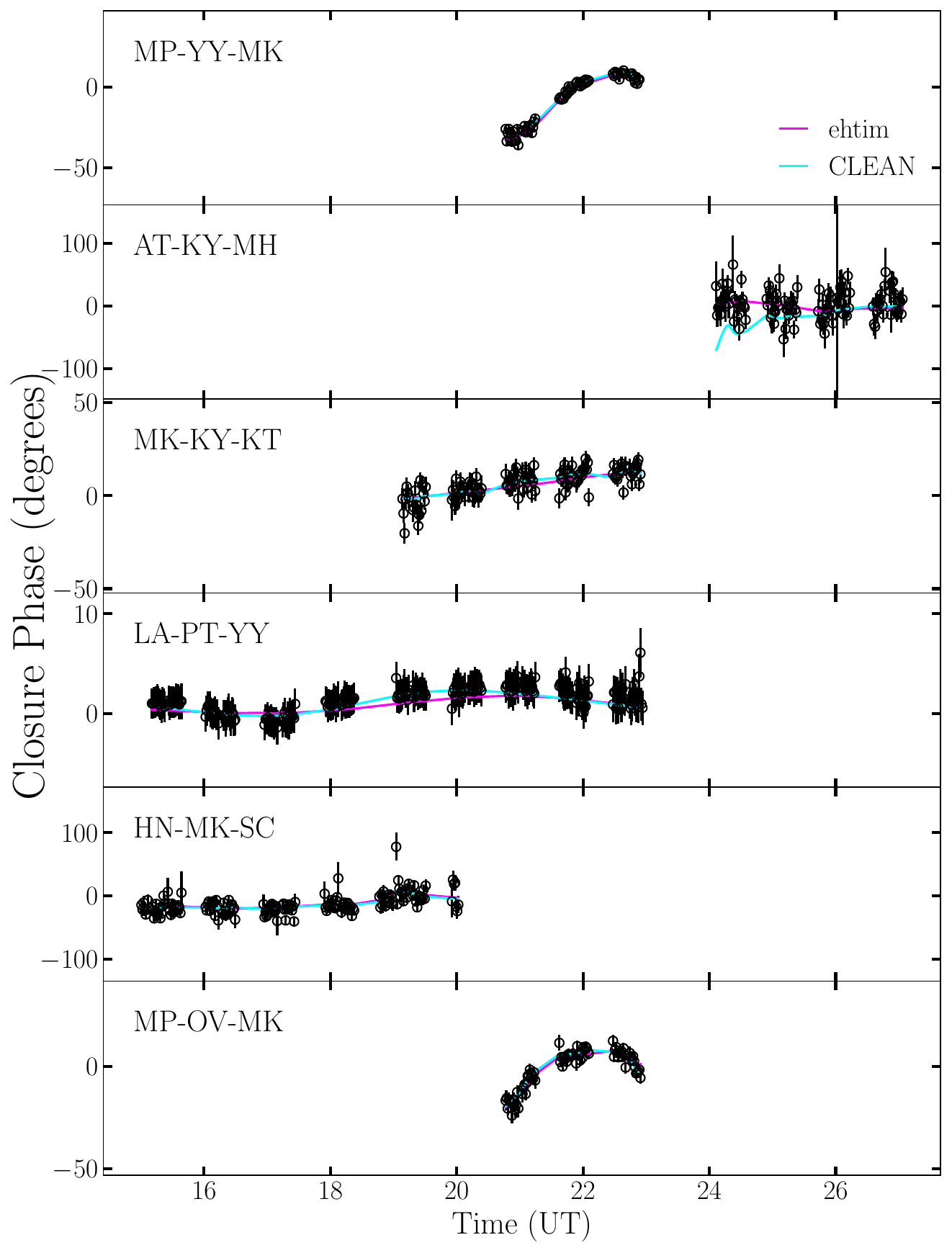}
\caption{Closure phase as a function of time for six selected triangles: observed data (black), RML model (magenta), and CLEAN model (cyan). Both models generally align with the observed closure phases. However, the RML model is particularly effective in explaining the closure phases of triangles formed with very long baselines (e.g., AT-KY-MH), while the CLEAN model tends to better explain those formed with short baselines (e.g., LA-PT-YY). Refer to Table~\ref{tab:antennas} for the stations corresponding to each station code.}

\label{fig:cphase}
\end{figure}

\section{Analysis \& Results}
\label{sec:analysis}

\subsection{Limb-brightened Jet Morphology}
\label{sec:analysis:morphology}

In Figure~\ref{fig:fiducial}, we present images from CLEAN (middle) and RML (right) and compare them with the VLBA-only image at 22 GHz observed on January 5, 2020 (Project Code: BP243, \citetalias{Park2021b}) shown on the left. The VLBA-only image exhibits a central ridge-brightened jet morphology. In contrast, the GVA images reveal a limb-brightened jet morphology at the same scale. The limb-brightened jet feature is more pronounced in the RML image, possibly because the method can more effectively capture fine-scale jet structures than CLEAN. This is attributed to its sensitivity in detecting a smoother brightness distribution and its capability to achieve super-resolution, as explained in Appendix~\ref{appendix:observations}.

We present the closure phases of six selected triangles of the data and models in Figure~\ref{fig:cphase}. In general, both CLEAN and RML models fit the data well, as expected from the reduced chi-squares for the closure phase close to unity ($\chi^2_{\rm CP}=$ 1.23 and 1.20 for the CLEAN and RML models, respectively; see Figure~\ref{fig:fiducial}). However, the RML model fits the triangle consisting of very long baselines (e.g., AT-KY-MH) better than the CLEAN model, while vice versa for the triangle consisting of short baselines (e.g., LA-PT-YY). This result indicates that the RML model better represents the fine-scale source structure than the CLEAN model. On the contrary, the RML model is less sensitive to the extended jet structure, as the closure phases of short-baseline triangles have small deviations from zero due to the large error bars inflated by adding the systematic uncertainty (which is used for RML imaging), while CLEAN used the complex visibilities without adding any systematic uncertainty. This inference can be observed in the images (Figure~\ref{fig:fiducial}).

\begin{figure*}[t]
\centering
\includegraphics[width=0.51\linewidth]{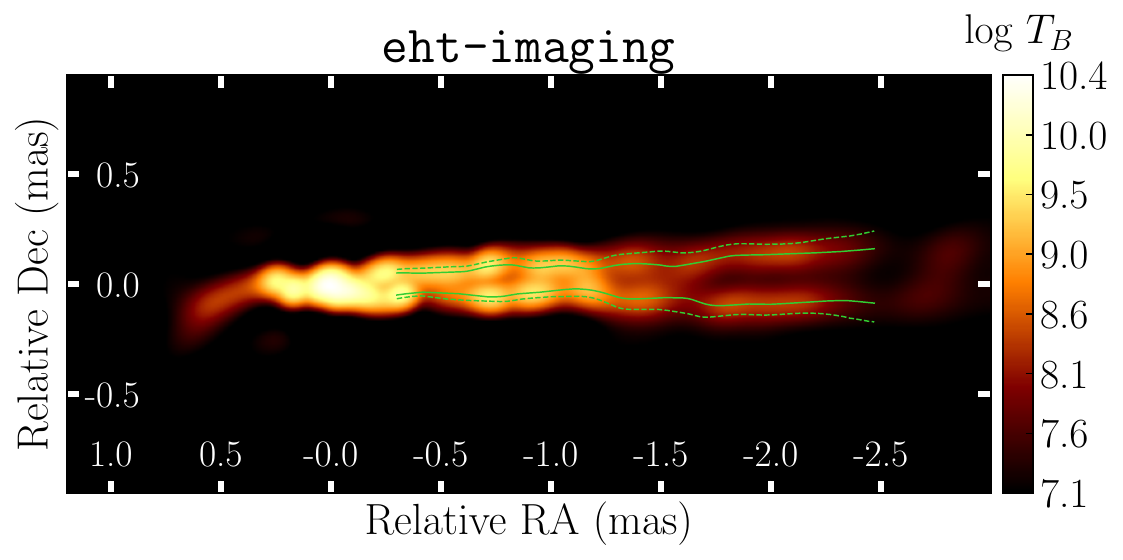}
\includegraphics[width=0.47\linewidth]{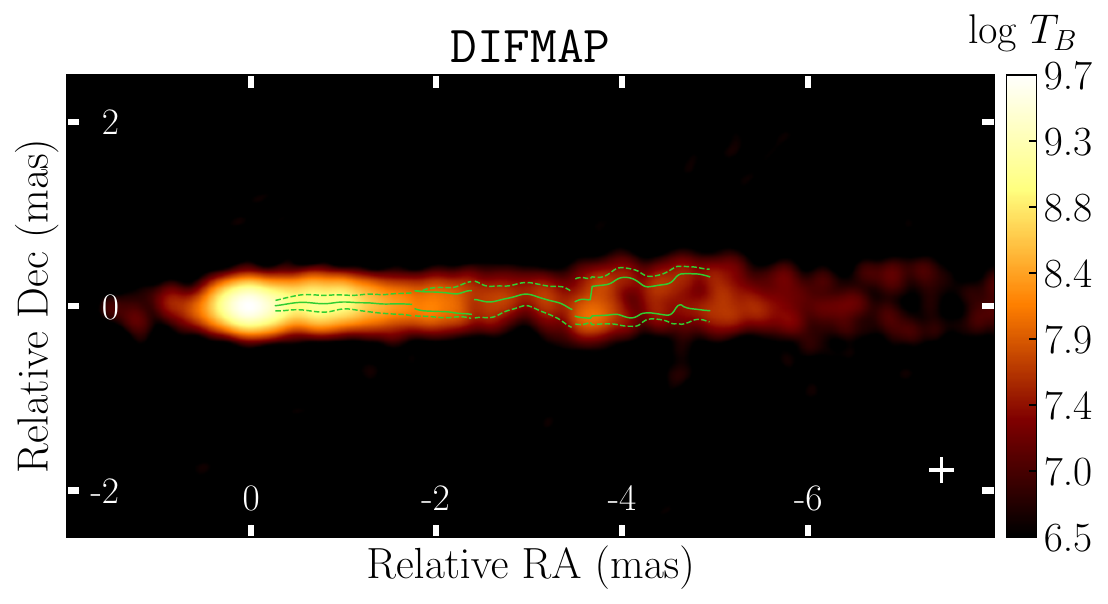}
\caption{RML (left) and CLEAN (right) images of the NGC 315 jets observed with the GVA. Both images are rotated clockwise by $40^\circ$. Note that the field of view of the RML image is smaller than that in Figure~\ref{fig:fiducial}, as we did not attempt to measure the jet width at distances greater than $\sim2.5$ mas from the core, due to the faintness of the reconstructed jet structure. The green solid lines represent the location of jet ridge(s), and the green dashed lines indicate the location of the jet edge(s) (see the text for more details). The CLEAN image is obtained by restoring the CLEAN models with a circular beam, with a size of the major axis FWHM of the synthesized beam, approximately $\sim0.25$ mas.}
\label{fig:rotate}
\end{figure*}

The counterjet emission is detected in both CLEAN and RML images (see also Figure~\ref{fig:rotate}). In the RML image, the counterjet emission is bright and appears symmetric with respect to the approaching jet up to a distance of $\sim0.3$ mas from the core. Beyond this distance, the counterjet rapidly becomes fainter, and limb-brightening is not detected, despite its presence in the approaching jet within the same distance range. It is unclear whether limb-brightening is absent in the counterjet, or if only some parts of the counterjet emission in this region are imaged due to the faintness and limited sensitivity of the data. We computed the expected intensity level of the counterjet emission by employing the mean intensity of the approaching jet in the distance range of 0.5–1.5 mas from the core, and using the apparent speed of $\beta_{\rm app} = 0.75\pm0.20$ and the spectral index of $\alpha = -0.81 \pm 0.40$ in the distance range, as reported in \citetalias{Park2021b}. The logarithm of the expected brightness temperature on the counterjet side in the same distance range is $7.80^{+0.48}_{-0.63}$ (90\% confidence interval). This calculation explains the non-detection of the counterjet beyond $\sim0.5$ mas.

In our previous study \citepalias{Park2021b}, we reported indications of a limb-brightened morphology in the NGC 315 jet at 43 GHz using the HSA. However, this feature was discernible only in a super-resolved image. Our GVA image, benefiting from the high angular resolution achieved by very long baselines, particularly the LBA baselines, and including numerous telescopes and high-sensitivity stations (e.g., phased VLA, phased ATCA, and the Effelsberg 100 m telescope), has revealed unambiguous evidence of a limb-brightened jet in NGC 315 for the first time. We conducted imaging reconstructions using only parts of the GVA data in Appendix~\ref{appendix:jackknife} and found no clear evidence of limb-brightening in the jets. This result indicates that the limb-brightened nature of the jet can only be robustly revealed with the full utilization of the GVA.

\subsection{Jet Collimation Profile}
\label{sec:analysis:collimation}

We derived the jet radius\footnote{The jet radius is assumed to be half the jet width.} as a function of jet distance following the procedure outlined in \citetalias{Park2021b}. Specifically, for the CLEAN image, we restored the CLEAN model using a circular beam with the size of the major axis of the synthesized beam ($\sim0.25$ mas) to eliminate the impact of the restoring beam straightforwardly. In contrast, the RML image was blurred using a circular Gaussian kernel with a size corresponding to 30\% of the diffraction limit, which is $\approx67 \mu{\rm as}$ (see Appendix~\ref{appendix:survey}). 

We applied a clockwise rotation of $40^\circ$ to align the jet axis with the x-axis of the maps (Figure~\ref{fig:rotate}). At each jet distance, we obtained a transverse intensity profile (along the y-axis of the rotated maps) and fitted a single or double Gaussian function to the profile, depending on the number of peaks identified. If the profile is fitted with a single Gaussian function, which is the case for the CLEAN image in certain distance ranges, we obtained the intrinsic jet width by subtracting the restoring beam FWHM from the measured jet FWHM in quadrature. If the profile is fitted with a double Gaussian function, then the width is defined as the distance between the outer edges\footnote{The edge is defined to be half of the FWHM separated from the peak of the Gaussian. We used the beam-subtracted FWHM.} of the two Gaussian functions, following the approach in previous studies of the jet width measurement for M87 \citep[e.g.,][]{AN2012, Hada2013}. 

The source structure near the core is compact, and the determination of the jet width in this region may not be robust. Therefore, we determined the jet widths at distances beyond the major axis beam size from the core. While we initially derived the intrinsic jet width at each distance bin (with a size corresponding to the image pixel size), many of these measurements are not independent due to the finite beam size. Consequently, we binned the jet widths in distance with a bin size equivalent to half the major-axis beam size, taking the median value of the widths in each bin as a representative jet width and assuming 1/10 of the major-axis beam size for the uncertainty of the width. 

\begin{figure}[t]
\centering
\includegraphics[width=\linewidth]{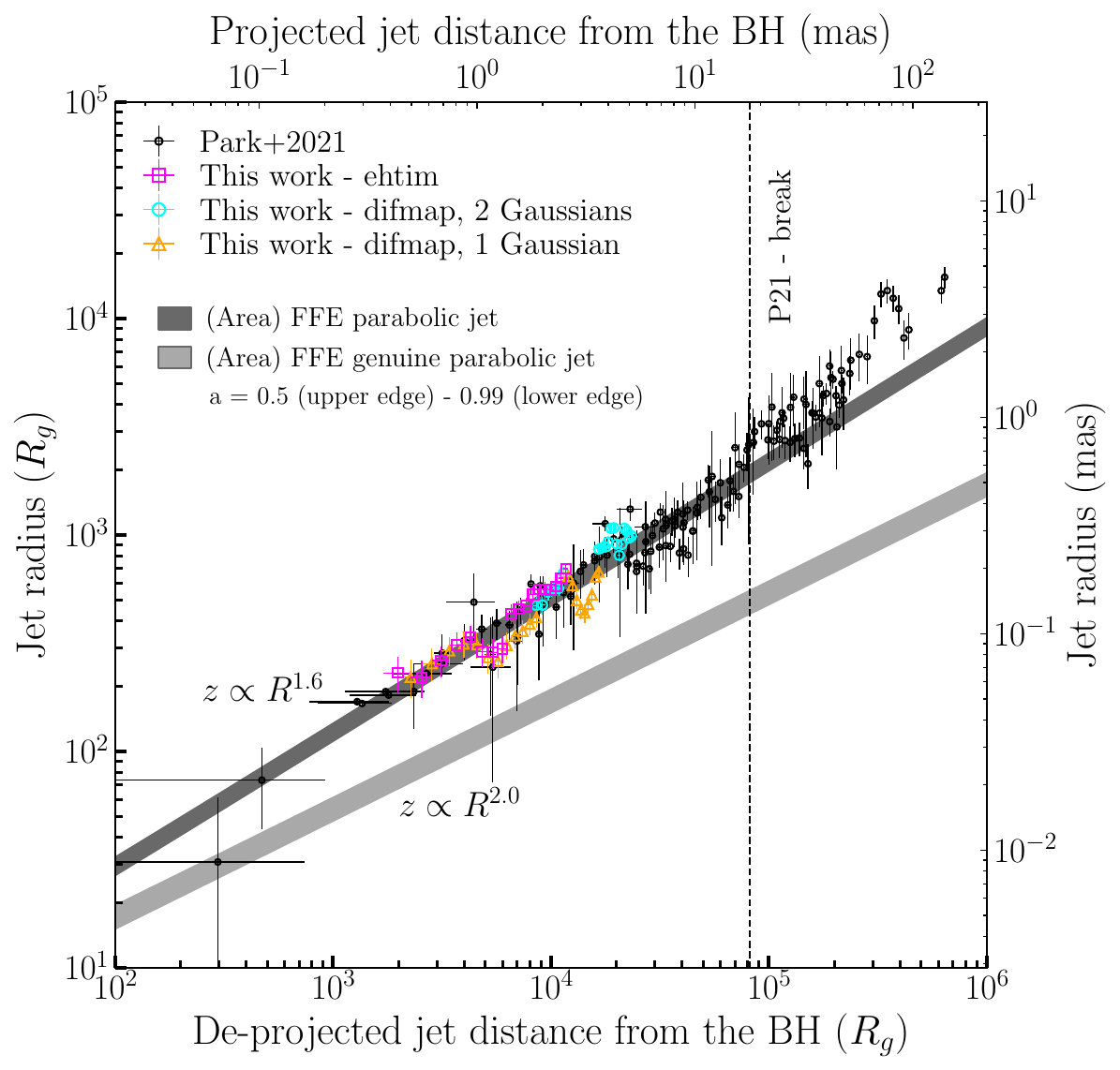}
\caption{The figure displays the jet radius as a function of de-projected distance from the black hole in units of $R_g$, with data from the RML image (magenta), CLEAN image (from 2 Gaussian fitting -- cyan, from 1 Gaussian fitting -- orange), and the previous study \citetalias{Park2021b}. The vertical black dashed line denotes the location of the jet collimation break. In the figure, the dark-gray (light-gray) segment denotes the outermost poloidal field line anchored to the equator of the event horizon, derived from the force-free electrodynamic (FFE) solution for $\kappa = 0.75$ ($\kappa = 1$) over a black hole spin range of 0.5--0.99. This corresponds to $z \propto R^{1.6}$ ($z \propto R^{2}$) asymptotically, where $R$ and $z$ denote the jet radius and distance, respectively. Here, $\kappa$ is a power-law index on the radius in the approximate formula of the magnetic stream function in polar coordinates in the Boyer-Lindquist frame \citep{Tchekhovskoy2010, Nakamura2018}.}
\label{fig:width}
\end{figure}

We present the jet radius as a function of de-projected\footnote{In our de-projection procedure, we adopted the black hole mass measurement of $M_{\rm BH} = 2.08 \times 10^9 M_\odot$, as recently reported by observations from the Atacama Large Millimeter/submillimeter Array (ALMA) \citep{Boizelle2021}, which detected CO emissions within the black hole's sphere of influence. The black hole mass we used contrasts with the estimate of $1.6 \times 10^9 M_\odot$ from our previous work \citepalias{Park2021b}, derived from the $M_{\rm BH}-\sigma_v$ relation. We also used a jet viewing angle of $49.8^\circ$, as measured by modeling the kpc-scale jet and counterjet \citep{LB2014}.} jet distance from the black hole in units of $R_g$ in Figure~\ref{fig:width}. To convert the jet distance from the core into the distance from the black hole, we corrected for the core-shift effect using the position of the 22 GHz core relative to the black hole, which was determined to be $\sim0.11\pm0.06$ mas in \citetalias{Park2021b}. It is possible that an additional apparent core shift exists between the RML and CLEAN images, potentially attributed to variations in the sizes of the blurring kernels. We have corrected for this shift, estimated to be $\sim0.063$ mas, by assuming that the brightest pixel in the RML image aligns with the core at 22 GHz when measuring the jet width. The jet collimation profile derived from the GVA images is in good agreement with the profile obtained in our previous study \citepalias{Park2021b}. We could not detect a significant jet collimation break at the apparent distance of $\sim1$ mas from the core as claimed in another study \citep{Boccardi2021}, and we discuss it in more detail in Appendix~\ref{appendix:boccardi}.

It is noted that the jet width does not monotonically increase with distance, but exhibits local fluctuations. This behavior can also be identified in the GVA images (Figure~\ref{fig:fiducial}). The fluctuations exist not only in the representative RML image but also in other RML images showing the lowest overall closure $\chi^2$ values (Appendix~\ref{appendix:survey}), and thus can be considered a real jet feature. These fluctuations may manifest as recollimation and re-expansion features, which can be naturally produced due to the mismatch between the jet and the ambient medium \citep[e.g.,][]{KF1997, Agudo2001, Mizuno2015, Fuentes2018}. However, we did not observe a significant increase in jet intensity at the locations of local jet contraction, which is expected for recollimation shocks \citep[e.g.,][]{BT2018}. Another possibility is that the fluctuations may be caused by the dynamic interaction of the jet with the outer ambient medium, which is often seen in numerical simulations of LLAGNs \citep[e.g.,][]{Tchekhovskoy2011, Chatterjee2019, Fromm2022}.

\subsection{Constraints on Jet Spine Emission}
\label{sec:analysis:spine}

In our GVA image of NGC 315 (Figure~\ref{fig:fiducial}), the jet structure is characterized by two limbs, with no clear evidence of a central ridge emission. We refer to the dim, on-axis region of the jet as the "spine" and to the limbs as the "sheath" (see Section~\ref{sec:discussion} for discussion). We attempt to infer the intensity level of the potential spine emission, which could be useful in discussing its speed and the transverse velocity structure of the jet.

We analyzed the intensity ratio between spine and sheath emissions using the representative RML image (Figure~\ref{fig:fiducial}, right). We extracted the transverse intensity profiles of the jet within the 0.5–2.35 mas distance range, where the two limbs are distinctly resolved. Although the limbs are resolved beyond this distance range, the overall intensity does not significantly exceed the noise level of the image; thus, we disregarded this outer range from our analysis. Subsequently, we computed the residuals by comparing the observed intensity profiles with the best-fit double Gaussian functions at each distance bin. We found that the residual intensity in the spine, defined as the midpoint between the two fitted Gaussians, is always smaller than approximately two times the root-mean-square (rms) of the residual intensity profile at each distance bin. Therefore, we conclude that the spine emission is not significant in the RML image. We constrained an upper limit on the spine emission by deriving three times the rms of the residual intensity profile divided by the mean intensity of the two fitted Gaussians (See Section~\ref{sec:discussion:Doppler} for more details).

\section{Discussion}
\label{sec:discussion}

The phenomenon of limb brightening in certain AGN jets remains a topic of active debate. The velocity gradient across the jet, featuring a faster-moving core or "spine" along the axis enveloped by a slower "sheath," has been central to an established explanation based on previous studies \citep[e.g.,][]{Sol1989, Komissarov1990, Laing1999, Perlman1999, Chiaberge2000, Ghisellini2005, Hardee2007, Mizuno2007, TG2008, Clausen-Brown2013}. This setup is commonly referred to as the "faster-spine and slower-sheath" model, or simply the "spine-sheath" model, and has been suggested by various studies. The transverse velocity gradient could be associated with different jet streamlines originating in various parts of the black hole magnetosphere and the accretion disk \citep[e.g.,][]{Komissarov2007, Tchekhovskoy2008}, or with the interaction between the relativistic jet and the ambient medium \citep[e.g.,][]{Chatterjee2019}.

According to this model, when observing nearly head-on jets, such as in the case of blazars, the emission from the faster jet spine appears more dominant due to the enhanced Doppler boosting effects compared to the slower sheath. Conversely, for jets observed from larger angles typical of radio galaxies, brighter emission originates from the slower sheath, resulting in limb brightening. This model has been proposed to account for observed limb brightening in sources like M87 \citep[e.g.,][]{Mertens2016, Walker2018}, 3C 84 \citep[e.g.,][]{Nagai2014, Giovannini2018}, Mrk 501 \citep[e.g.,][]{Giroletti2004}, Cygnus A \citep[e.g.,][]{Boccardi2016b, Boccardi2016a}, and Centaurus A \citep{Janssen2021}.

Confusion often arises from the inconsistent use of the terms "spine" and "sheath" in the literature. We aim to clarify definitions based on General Relativistic Magnetohydrodynamic (GRMHD) simulations (see \citealt{Mizuno2022} for a review) of hot accretion flows. It is believed that LLAGNs are powered by hot accretion flows \citep[e.g.,][]{YN2014}, and NGC 315, with a mass accretion rate inferred to be $\approx10^{-4}\dot{M}_{\rm Edd}$ \citep{Gu2007, Worrall2007, Morganti2009, Ricci2022}, is classified as a LLAGN.

A relativistic jet, believed to originate from a spinning black hole, is an outflow of highly magnetized material potentially accelerated to relativistic speeds depending on magnetization and energy conversion efficiency \citep[e.g.,][]{Komissarov2007, TT2013, Kino2022}. Surrounded by a geometrically thick disk, the jet along large-scale magnetic field lines is enclosed within a funnel region for a hot accretion flow system. The jet region can be subdivided into two areas: the on-axis region along the jet funnel, and the jet edge, which roughly coincides with the outermost large-scale magnetic field. This field is approximately anchored to the equatorial plane of the black hole’s event horizon \citep[e.g.,][]{Nakamura2018}. In jets launched by spinning black holes, the edges experience more efficient bulk jet acceleration than the axis, a fundamental characteristic of MHD jets from such sources \citep[e.g.,][]{Komissarov2007, Komissarov2009, Tchekhovskoy2008, Tchekhovskoy2009, Tchekhovskoy2010, Penna2013, Takahashi2018, PT2020, Kino2022, Yang2024, Ricci2024}. 

In this paper, we define the "jet spine" as the on-axis region and the "jet sheath" as the edge region, acknowledging that others have used divergent definitions \citep[e.g.,][]{Fromm2022}. Notably, the continuous distribution of physical quantities suggests that defining the outflow as two distinct regions may be an oversimplification.

Outside the relativistic jet, it is believed that a wind, also known as a disk-wind, surrounds the jet. This wind is a moderately magnetized outflow with a high mass flux \citep{Sadowski2013, Yuan2015, Yuan2022, Nakamura2018, Yang2021} and is thought to originate from the accretion flows. It remains subrelativistic due to its higher mass loading. The jet’s parabolic shape is attributed to being collimated by the wind pressure\footnote{Note that the transition of the jet shape from a parabolic to a conical/hyperbolic shape may be associated with a change in the pressure profile of the external medium \citep[e.g.,][]{AN2012} and/or a transition from a magnetically dominated jet to an equipartition regime \citep[e.g.,][]{Beskin2017, Nokhrina2019, Nokhrina2020, Kovalev2020}.}, which itself remains uncollimated and exhibits a broad opening angle \citep[e.g.,][]{Sadowski2013, YN2014, Park2019a}.

It is noted that this picture of relativistic jets collimated by pressure-driven winds launched from hot accretion flows is based on previous theoretical and observational studies of LLAGNs. Depending on the black hole inflow-outflow system, AGN jets can also be collimated by other types of ambient medium, such as dense and cold gas clouds \citep[e.g.,][]{Kino2021, Park2024}, hot mini-cocoons surrounding the jets \citep[e.g.,][]{Savolainen2023}, and radiation-driven winds and strong radiation pressure from the accretion disk \citep[e.g.,][]{Ohsuga2009, OM2011, Hada2018}.

Earlier studies linking limb brightening in AGN jets to the fast-spine and slow-sheath model often did not specify  their definitions of spine and sheath clearly. Typically, they interpreted limb brightening as emission from the slower sheath, attributing the central dimness to the faster, more strongly beamed spine region \citep[e.g.,][]{Giroletti2004, Nagai2014, Boccardi2016a, Mertens2016, Kim2018, Giovannini2018, Walker2018, Janssen2021}. This rationale also considered that the sheath's reduced speed might result from the jet's interaction with the surrounding medium. Consequently, their concepts of spine and sheath likely correspond to our on-axis and edge regions\footnote{If previous studies interpreted the observed two limbs as slower winds surrounding the jet, it would be difficult to argue that we are observing the jets in those sources, since the winds are uncollimated and subrelativistic.}.

\subsection{Insights from GRMHD Jet Models Applied to M87: A Slow-Spine and Fast-Sheath Jet Structure}

In contrast to the faster-spine and slower-sheath structure mentioned above, GRMHD simulations for black hole hot accretion flow systems show a slow-spine, fast-sheath structure of the relativistic jet along large-scale magnetic fields \citep[e.g.,][]{McKinney2006}. \cite{Nakamura2018} show that the observed collimation profile of the M87 jet closely aligns with the boundary between the jet and the surrounding winds. This finding is further confirmed by subsequent GRMHD simulations at scales down to $\approx 10^5 \ R_g$ \citep{Chatterjee2019}. These results indicate that the north and south limbs observed in the M87 jet may trace the jet boundary. Thus, \cite{Nakamura2018} concluded that the two limbs of the M87 jet represent a faster jet sheath that dominates the spine emission due to strong Doppler boosting effects.

\begin{figure}[t]
\centering
\includegraphics[width=\linewidth]{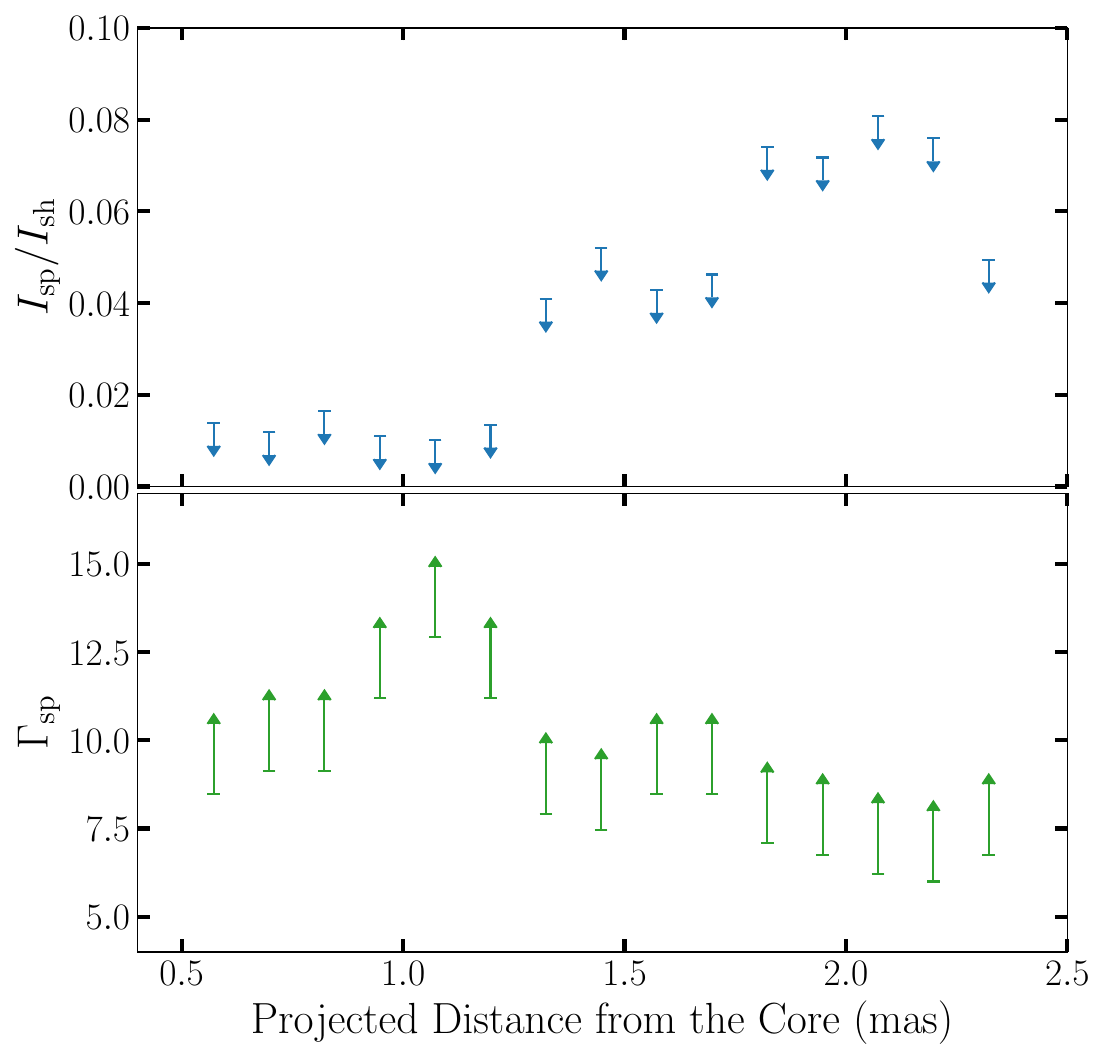}
\caption{\emph{Top:} Upper limit on $I_{\rm sp} / I_{\rm sh}$ as a function of projected distance from the core, as derived in Section~\ref{sec:analysis:spine}. \emph{Bottom:} Lower limit on $\Gamma_{\rm sp}$ required to explain the upper limit on $I_{\rm sp} / I_{\rm sh}$ using the differential Doppler boosting effect.}
\label{fig:doppler}
\end{figure}

Nonetheless, our new GVA image of NGC 315 reveals a limb-brightened jet morphology that challenges the previously suggested slow-spine and fast-sheath jet model, which assumes uniform emissivity across the jet. The jet’s viewing angle for NGC 315 is $\approx50^\circ$, substantially larger than the $\approx17^\circ$ viewing angle of the M87 jet. Furthermore, the jet collimation profile of NGC 315 appears to be analogous to that of M87 (\citetalias{Park2021b}, \citealt{Boccardi2021}), indicating that the pronounced dual limbs seen in the NGC 315 image likely demarcate the interface between the jet and the wind\footnote{Both M87 and NGC 315 are believed to be LLAGNs, and a similar interpretation based on GRMHD simulations of hot accretion flows may be applicable.}. Assuming uniform emissivity within the jet and a slow-spine and fast-sheath configuration, one would anticipate more enhanced spine emission at NGC 315's large viewing angle, with sheath emission mostly beamed away. However, our observations clearly exhibit limb brightening, suggesting that the simple slow-spine and fast-sheath framework, assuming uniform synchrotron emissivity, may not be universally valid, especially in the context of the NGC 315 jet.

\subsection{Could the Limb-Brightening of NGC 315's Jet Be Due to the differential Doppler Effect?}
\label{sec:discussion:Doppler}

It is illustrative to examine the required speeds if the limb brightening of NGC 315 is associated with different velocities at the jet spine and the sheath. If we assume uniform emissivity across the jet, we can expect that intrinsic spine emission ($I_{\mathrm{sp}}$) can be Doppler de-boosted in comparison to intrinsic sheath emission ($I_{\mathrm{sh}}$) by a factor expressed as:
\begin{equation}
\frac{I_{\mathrm{sp}}}{I_{\mathrm{sh}}} \equiv \left(\frac{\delta_{\mathrm{sp}}}{\delta_{\mathrm{sh}}} \right )^{2-\alpha} = \left [ \frac{\Gamma_{\mathrm{sh}}(1 - \beta_{\mathrm{sh}}\cos\theta)}{\Gamma_{\mathrm{sp}}(1 - \beta_{\mathrm{sp}}\cos\theta)} \right ]^{2-\alpha},
\label{eq:doppler}
\end{equation}
where $\theta$ denotes the jet viewing angle assumed to be the same for the spine and sheath emission, $\delta_{\mathrm{sp}}$ and $\delta_{\mathrm{sh}}$ denote the Doppler factors for the spine and sheath emissions, respectively, $\Gamma_{\mathrm{sp}}$ and $\Gamma_{\mathrm{sh}}$ represent the corresponding bulk Lorentz factors for the spine and sheath, $\beta_{\mathrm{sp}}$ and $\beta_{\mathrm{sh}}$ represent the corresponding intrinsic jet speeds in units of the speed of light for the spine and sheath, and $\alpha$ is the spectral index defined as $I_\nu \propto \nu^\alpha$ \citep{Walker2018}.

In Section~\ref{sec:analysis:spine}, we derived upper limits on $I_{\mathrm{sp}}/I_{\mathrm{sh}}$ in the selected distance range of 0.5--2.35 mas (corresponding to a de-projected distance range of $(2.3-10.8) \times 10^3\ R_g$), where the two limbs are clearly resolved. We derived lower limits on the Lorentz factor of the putative spine emission to account for the upper limits on the intensity ratio between the jet spine and sheath, invoking the differential Doppler boosting effect, using Equation~\ref{eq:doppler}. We used linear interpolation of the jet intrinsic speed and spectral index values derived in \citetalias{Park2021b} for $\beta_{\rm sh}$ and $\alpha_{\rm sh}$, respectively. Figure~\ref{fig:doppler} presents the upper limit on $I_{\mathrm{sp}}/I_{\mathrm{sh}}$ (top) and lower limit on $\Gamma_{\rm sp}$ (bottom) as a function of projected distance from the core. The inferred lower limit on $\Gamma_{\rm sp}$ is in the range of $\approx6.0$--12.9. We note that $\Gamma_{\rm sh}$ used for deriving the lower limit on $\Gamma_{\rm sp}$ is in the range of $\approx1.4$--2.9.

However, as mentioned earlier (Section~\ref{sec:discussion}), it is challenging to accelerate the on-axis region of the jet more efficiently than the jet's edge regions, according to MHD models and simulations. Previous observations have revealed that jet collimation and acceleration occur simultaneously at distances up to $\approx 10^4-10^5\ R_g$ from the black hole (\citetalias{Park2021b}, \citealt{Boccardi2021, Ricci2022}), which implies that the jets are accelerated through an MHD process \citep[e.g.,][]{Komissarov2007, Lyubarsky2009, Nakamura2018}. Therefore, reconciling this result with the inferred relativistic jet spine with $\Gamma_{\rm sp} > 6.0-12.9$, would be challenging, as the MHD model predicts the opposite. Furthermore, the limb-brightening observed in the Cygnus A jet, for which the jet viewing angle is thought to be even larger ($\sim74.5^\circ$; \citealt{Boccardi2016a}) than that of the NGC 315 jet ($\sim50^\circ$; \citealt{LB2014}, \citetalias{Park2021b}), is even more challenging to explain with the fast-spine and slow-sheath jet velocity configuration.

\subsection{Higher Emissivity at the Edges of the NGC 315 Jet}

An alternative explanation suggests that the jet's edges have intrinsically higher synchrotron emissivity than its central spine. Using a steady, axisymmetric force-free jet model, \cite{Takahashi2018} demonstrated that the limb-brightened morphology observed in the M87 jet at de-projected distances $\lesssim 1,000\ R_g$ \citep[e.g.,][]{Hada2016, Walker2018, Kim2019, Park2021c, Lu2023}, can be produced when the non-thermal particles are distributed in a hollow paraboloidal shape\footnote{In \cite{Takahashi2018}, an additional factor of $1/(1 - \beta \mu)$ was erroneously introduced in their equations (13) and (14), which calculate the observed intensity. Although this factor was explained as a retardation effect in their paper, it is actually not necessary in their model, where a paraboloidal jet is filled with steady continuous flows. We have confirmed that the conclusions in \cite{Takahashi2018} remain the same even after correcting the error.}. Additionally, they showed that a slow toroidal velocity field is necessary to produce a symmetric limb-brightened image, as it suppresses the relativistic beaming/debeaming effects on each side of the jet axis. Moreover, the pronounced brightness along the edges corresponds to regions where nonthermal electrons, responsible for synchrotron emission, are preferentially located. It is noteworthy that the width of the observed limbs may provide insights into the effective thickness of the jet sheath on which the non-thermal electrons are distributed \citep{Takahashi2018, Ogihara2019}. 

\cite{Bruni2021} presented Space-VLBI RadioAstron observations of the flat-spectrum radio quasar 3C 273, revealing a limb-brightened jet morphology at 1.6 GHz but a central ridge-brightened morphology at 4.8 GHz. It was suggested that these results are challenging to explain with only a transversely stratified jet velocity and require stratification in the emitting electron energy distribution.

The increased emissivity could be due to (i) more efficient particle acceleration at the jet edges and/or (ii) more effective mass loading onto the edges due to magnetic reconnection in the equatorial region of a magnetically arrested disk (MAD; \citealt{Igumenshchev2003, Narayan2003, Tchekhovskoy2011}) near the black hole horizon. It has been suggested that jet instabilities, such as the pinch instability, may be associated with particle acceleration in the jet \citep[e.g.,][]{McKinney2006, Nakamura2018}. Recent GRMHD\footnote{It is noteworthy that accurately computing the expected jet emission based on GRMHD simulations is challenging due to the difficulty in handling low-density regions \citep[e.g.,][]{Fromm2022}. These simulations usually employ a 'floor' procedure that resets the density if it falls below a minimum value \citep[e.g.,][]{EHT2019e}. \cite{Chael2024} has proposed a novel hybrid approach that jointly uses GRMHD and general relativistic force-free electrodynamics (FFE) simulations to address this issue.} simulations have demonstrated that nonthermal electrons are essential to explain the observed limb brightening in AGN jets \citep{Cruz-Osorio2022, Fromm2022, Yang2024}. They have shown that nonthermal electrons, accelerated through magnetic reconnection in the jet, are necessary to explain the broadband spectral energy distribution of M87 \citep{Cruz-Osorio2022, Fromm2022} as well as the observed jet collimation profile \citep{Cruz-Osorio2022, Fromm2022, Yang2024}. \cite{Wang2024} have suggested that stochastic Fermi-type particle accelerations in turbulent shearing flows, i.e., classical Fermi II and gradual shear acceleration \citep[e.g.,][]{Rieger2019}, are necessary to explain the limb-brightened jet morphology of M87 over a long distance range, down to the distance of $\approx4\times10^5\ R_g$ \citep[e.g.,][]{Cheung2007, Park2019b}, as well as the high-energy emission observed in the inner jet of M87. A recent study of NGC 315 also indicates a possibility of energy loss at the boundary between the jet and the surrounding medium (Kino, M., et al. Submitted).

The process of efficient mass loading into the jet boundary has been demonstrated in recent GRMHD simulations. These simulations indicate that magnetic reconnection occurring on the equatorial plane of the accretion disk near the black hole can generate MeV $\gamma$-ray photons. This occurs due to antiparallel magnetic fields that are compressed into a thin layer in the MAD \citep{Porth2021, Ripperda2022, Kimura2022, Chen2023}. The resulting photons then produce electron-positron (\(e^+e^-\)) pairs through interactions. These pairs are subsequently loaded onto the jet, particularly near its edges, following the magnetic field lines anchored to the equatorial plane of the black hole's event horizon. As a result, the on-axis region of the jet would contain fewer particles, leading to negligible spine emission irrespective of the jet's transverse velocity structure. \cite{Ricci2022}, employing the model proposed by \cite{Nokhrina2019, Nokhrina2020} and the size of the jet acceleration and collimation zone (\citetalias{Park2021b}, \citealt{Boccardi2021, Ricci2022}), have proposed that NGC 315 may be in a MAD state (see also \citealt{NP2024} for related discussion), suggesting that the higher mass loading explanation could be viable.

\section{Summary \& Conclusions}
\label{sec:conclusion}

In this paper, we present the observation of NGC 315 conducted with the GVA at 22 GHz. The observation utilized a total of 22 telescopes across the United States, Europe, East Asia, South Africa, and Australia, achieving extensive $(u,v)$-coverage. Using both inverse modeling with the CLEAN method and forward modeling with the RML method, we have clearly resolved the NGC 315 jet transversely at parsec scales for the first time. The jet, known for decades to exhibit a central ridge-brightened morphology, now reveals a limb-brightened morphology in our new observations, thanks to the high-quality VLBI data obtained by the GVA. This finding confirms the hint of limb-brightening in the parsec-scale jet observed in the image convolved with a super-resolution beam from the HSA data at 43 GHz in our previous study \citepalias{Park2021b}.

Having resolved the jet transversely, we robustly derived a jet collimation profile within the de-projected jet distance range of approximately $2,000 - 23,000\ R_g$. This profile is consistent with that derived in our previous study \citepalias{Park2021b}. We found no clear evidence of a jet collimation break at the location suggested by another study \citep{Boccardi2021}, indicating that discrepancies between \citetalias{Park2021b} and \cite{Boccardi2021} may stem from differences in data, analysis methods, and potential source variability.

We explore several models to explain the observed limb-brightening in the NGC 315 jet. Firstly, we assess whether the slow-spine and fast-sheath jet model, as proposed by \cite{Nakamura2018} based on GRMHD simulations, FFE modeling, and various previous MHD studies of relativistic jets, is applicable to NGC 315. They demonstrated that the jet collimation profile observed in M87 aligns well with the boundary between the jet and the outer wind. Furthermore, MHD models typically generate a slower on-axis region and faster jet edge regions, due to more efficient bulk jet acceleration occurring at the edges through the magnetic nozzle effect, which produces the slow-spine and fast-sheath jet structure. This structure can naturally lead to limb-brightening in jets viewed at small angles due to more enhanced Doppler boosting at the jet edges compared to the on-axis region. This model was proposed to explain the limb brightening observed in the M87 jet, which is known to have a small jet viewing angle ($\approx17^\circ$; \citealt{Mertens2016, Walker2018}). However, according to this model, the NGC 315 jet, which is viewed at a much larger angle ($\approx50^\circ$; \citealt{LB2014}; \citetalias{Park2021b}) compared to the M87 jet, would be expected to exhibit a central ridge-brightened morphology, as the relativistic jet edge emission would be beamed away from our line of sight.

Secondly, the conventional fast-spine and slow-sheath jet model, assuming uniform synchrotron emissivity across the jet—often cited in limb-brightening studies of other AGN jets—may still apply to NGC 315. However, this model necessitates that the jet spine, if present, reaches speeds of $\Gamma_{\rm sp} > (6.0-12.9)$ in the de-projected jet distance range of $(2.3-10.8)\times10^3\ R_g$ to explain the non-detection of the jet spine in our GVA images. Current MHD models and simulations find it challenging to accelerate the on-axis region to such relativistic speeds at the distances discussed.

Lastly, we propose that the observed limb brightening may be linked to higher emissivity at the jet edges rather than a stratified jet velocity field. According to GRMHD simulations, increased emissivity could derive from (i) more efficient particle acceleration at the jet edges and/or (ii) more effective mass loading at the edges. The former might be related to pinch instability at the jet boundary \citep[e.g.,][]{McKinney2006, Nakamura2018} or magnetic reconnection within the jet \citep[e.g.,][]{Cruz-Osorio2022, Fromm2022, Yang2024}. The latter could be associated with magnetic reconnection in the innermost regions of the MAD \citep[e.g.,][]{Porth2021, Ripperda2022, Kimura2022, Chen2023}.

We suggest that future high-resolution imaging with EHT observations of NGC 315 will enable testing of the prevailing fast-spine and slow-sheath jet model, which assumes uniform emissivity. Previous observations have revealed the existence of a jet acceleration and collimation zone (\citetalias{Park2021b}; \citealt{Boccardi2021}; \citealt{Ricci2022}), a fundamental prediction of the MHD jet acceleration model \citep[see, e.g.,][]{Komissarov2007, Lyubarsky2009, Vlahakis2015, Nakamura2018}. Consequently, in this model, the jet spine—if present—would not be relativistic at short distances as it lacks the necessary length for significant acceleration. GRMHD simulations suggest jet speeds are $\Gamma \lesssim$ a few at distances up to several hundred $R_g$ from the black hole (see Figure 16 in \citealt{Nakamura2018}). Therefore, it is plausible that spine emission could be observable up to such distances. Beyond this region, the spine's increased velocity and the concentration of its emission along the jet axis—due to Doppler boosting—may render it undetectable from the observer's viewpoint (the NGC 315 jet is angled at $\approx50^\circ$ from our line of sight). It is noteworthy that if spectral index information of the spine exists, it may also constrain the cooling mechanism of nonthermal electrons in the jet, as suggested by \cite{Pu2017}.

Figure~\ref{fig:fiducial} convincingly demonstrates that the NGC 315 jet, previously known for its central ridge-brightened morphology, actually exhibits limb-brightening. This finding suggests the intrinsic structure of the NGC 315 jet is characterized by limb-brightening, a feature that remained elusive in earlier VLBI observations because of insufficient angular resolution. This result indicates the possibility that AGN jets, traditionally recognized for central ridge-brightening on pc scales, may in fact display limb-brightening, which went undetected in past studies due to limited angular resolution\footnote{However, dividing the jet morphology into central ridge-brightening and limb-brightening may be an oversimplification. For example, \cite{Fuentes2023} revealed a filamentary structure in the jet of the quasar 3C 279 by resolving the jet transversely using the space VLBI mission RadioAstron. The filamentary structure was interpreted as originating from plasma instabilities in a kinetically dominated flow. \cite{Nikonov2023} modeled the parsec-scale structure of the M87 jet with helical threads, which might be produced by the Kelvin-Helmholtz instability developed in the jet. Thus, the extent of the development of instabilities is another important factor in determining jet morphology. Instabilities can produce more complex jet geometries than a simple ridge or limb-brightened structure.}. Moreover, it emphasizes the challenge in investigating the origins of limb-brightening in sources where the transverse jet structures remain unresolved. To address this matter, we plan to conduct a dedicated statistical study of nearby AGN jets using high-resolution VLBI observations with GVA, the Global Millimeter VLBI Array (GMVA; \citealt{Zhao2022, KimDW2023}), the EHT, and the Black Hole Explorer (BHEX\footnote{\url{https://www.blackholeexplorer.org/}}; \citealt{BHEX2024}) space-VLBI mission in the future.

\begin{acknowledgements}

The authors appreciate the referee's constructive comments, which have improved the paper. J.P. expresses gratitude to Tuomas Savolainen for providing the \ehtim{} script, which served as the foundation for the RML imaging performed in this research. The authors express their thanks to Yuri Kovalev, the MPIfR internal reviewer, for carefully reading the manuscript and providing valuable comments that improved the paper. J.P. acknowledges ﬁnancial support through the EACOA Fellowship awarded by the East Asia Core Observatories Association, which consists of the Academia Sinica Institute of Astronomy and Astrophysics, the National Astronomical Observatory of Japan, Center for Astronomical Mega-Science, Chinese Academy of Sciences, and the Korea Astronomy and Space Science Institute. This work was supported by a grant from Kyung Hee University in 2023 (KHU-20233238). This work was supported by the BK21 FOUR program through National Research Foundation of Korea (NRF) under Ministry of Education (Kyung Hee University, Human Education Team for the Next Generation of Space Exploration) G.-Y.Z. acknowledges the support from the M2FINDERS project that is funded from the European Research Council (ERC) under the European Union's Horizon 2020 research and innovation programme (grant agreement No 101018682). This work supported by the MEXT/JSPS KAKENHI (Grant Numbers: JP24K07100, 23K03453, 21H04488, 21H01137, 22H00157). Y.M. is supported by the National Key R\&D Program of China (No. 2023YFE0101200), the National Natural Science Foundation of China (Grant No. 12273022), and the Shanghai Municipality orientation program of Basic Research for International Scientists (grant no. 22JC1410600). Scientific results from data presented in this publication are derived from the following EVN project code(s): GP060. The European VLBI Network is a joint facility of independent European, African, Asian, and North American radio astronomy institutes. The VLBA is an instrument of the National Radio Astronomy Observatory. The National Radio Astronomy Observatory is a facility of the National Science Foundation operated by Associated Universities, Inc. We are grateful to the staff of the KVN who helped to operate the array and to correlate the data. The KVN is a facility operated by the KASI (Korea Astronomy and Space Science Institute). The KVN observations and correlations are supported through the high-speed network connections among the KVN sites provided by the KREONET (Korea Research Environment Open NETwork), which is managed and operated by the KISTI (Korea Institute of Science and Technology Information). The Long Baseline Array is part of the Australia Telescope National Facility (\url{https://ror.org/05qajvd42}) which is funded by the Australian Government for operation as a National Facility managed by CSIRO.

\end{acknowledgements}

\begin{appendix} 
\label{appendix}

\section{Detailed Description of Observations and Data Reduction}
\label{appendix:observations}

On November 9, 2022, an observation of NGC 315 was carried out using the GVA at 22 GHz (Project Code: GP060). A total of 30 telescopes worldwide were initially scheduled for participation, including 10 VLBA stations (Station Codes: BR, FD, HN, KP, LA, MK, NL, OV, PT, SC), the phased VLA (YY), 10 EVN stations (EF, JB, O6, HH, MC, NT, TR, YS, MH, RO), 5 EAVN stations (KY, KU, KT, UR, T6), and 4 Long Baseline Array (LBA) stations (AT, MP, HO, CD). However, due to weather and technical challenges, two VLBA stations (BR, KP), two EVN stations (TR, RO), two EAVN stations (UR, T6), and two LBA stations (HO, CD) were unable to participate in the observation. The VLA and ATCA were phased up approximately every 15 minutes, as well as when the antennas moved more than $\sim15^\circ$ in the sky. The VLA was in the C configuration, and the ATCA was in the EW352 configuration with five participating antennas, each separated by less than approximately 350\,m from each other. The data were recorded in both right- and left-hand circular polarizations, using two-bit quantization in eight baseband channels, each with a bandwidth of 32 MHz. The recording rate was set at 2 Gbps, except for the LBA stations, whose recording rate was set at 1 Gbps, covering half of the frequency bandwidth. The correlation process was conducted at the Joint Institute for VLBI ERIC (JIVE) in the Netherlands \citep{Keimpema2015}. In Table~\ref{tab:antennas}, we present the basic properties of the antennas used in the GVA observation, including the mean SNR of the baselines associated with each station. The SNR is computed for Stokes I data averaged over one minute.

\begin{table*}[ht]
    \centering 
    \caption{Properties of the antennas used in the GVA observation \label{tab:antennas}}
    \begin{tabular}{cccccc}
        \hline
        Array & Station & Site Code & Longitude & Latitude & Mean SNR \\
        \hline 
        \multirow{8}{*}{VLBA} & Fort Davis       & FD & -103:56:41 &  30:38:06 & 38.8 \\
                               & Hancock          & HN &  -71:59:11 &  42:56:01 & 40.2 \\
                               & Los Alamos       & LA & -106:14:44 &  35:46:30 & 43.7 \\
                               & Mauna Kea        & MK & -155:27:19 &  19:48:05 & 41.7 \\
                               & North Liberty    & NL &  -91:34:26 &  41:46:17 & 35.7 \\
                               & Owens Valley     & OV & -118:16:37 &  37:13:54 & 42.3 \\
                               & Pie Town         & PT & -108:07:09 &  34:18:04 & 45.8 \\
                               & Saint Croix      & SC &  -64:35:01 &  17:45:24 & 15.2 \\
        \hline 
                    Phased VLA & Very Large Array & Y  & -107:37:06 &  34:04:44 & 59.6 \\
        \hline
          \multirow{8}{*}{EVN} & Effelsberg       & EF &  6:53:01   &  50:31:29 & 48.1 \\ 
                               & Hartebeesthoek   & HH & 27:41:07   & -25:53:23 &  9.2 \\
                               & Jodrell Bank     & JB &  -2:18:14  &  53:14:02 & 14.7 \\
                               & Noto             & NT & 14:59:20   &  36:52:34 & 16.9 \\
                               & Mets\"{a}hovi    & MH & 24:23:35   &  60:13:04 & 14.5 \\
                               & Medicina         & MC & 11:38:49   &  44:31:14 & 20.4 \\
                               & Onsala           & O6 & 11:55:34   &  57:23:45 & 19.7 \\
                               & Yebes            & YS & -3:05:12   &  40:31:29 & 28.6 \\
        \hline
         \multirow{3}{*}{EAVN} & KVN Tamna        & KT & 126:27:34  & 33:17:20  & 21.6 \\ 
                               & KVN Ulsan        & KU & 129:14:59  & 35:32:44  & 24.4 \\
                               & KVN Yonsei       & KY & 126:56:27  & 37:33:55  & 22.4 \\
        \hline
          \multirow{2}{*}{LBA} & ATCA             & AT & 149:33:53  & -30:18:46 & 26.9 \\ 
                               & Mopra            & MP & 149:05:58  & -31:16:04 & 20.5 \\
        \hline
    \end{tabular}
\tablecomments{Positive longitudes are east of the prime meridian.}
\end{table*}

We performed standard data reduction using the NRAO's Astronomical Image Processing System (AIPS; \citealt{Greisen2003}). It is noteworthy that the data of 3C 84, observed as a calibrator in the same observing run, have already been analyzed and published \citep{Park2024}. One noticeable difference from the reduction of the 3C 84 data is that we conducted global fringe fitting \citep{CS1983} with a solution interval of 1 minute across the entire frequency bandwidth, as opposed to the 10 seconds used for the 3C 84 data reduction. This adjustment was made because NGC 315 is much fainter than 3C 84. 

\subsection{Imaging and self-calibration}
\label{sec:observation:imaging}

We utilized two methods for imaging the data: inverse modeling with CLEAN \citep{Hogbom1974, Clark1980, RC2011}, implemented in the \difmap{} software package \citep{Shepherd1997}, and forward modeling with the regularized maximum likelihood (RML) method implemented in the \ehtim{} library\footnote{\url{https://github.com/achael/eht-imaging}} \citep{Chael2016, Chael2018, Chael2023}. The CLEAN algorithm is a widely used technique in radio interferometric imaging, employing an inverse modeling approach for sparse reconstruction in the image domain, relying on the point-source response of the interferometer, known as the dirty beam. The version of the CLEAN algorithm implemented in \difmap{} presupposes that the sky's brightness distribution is composed of discrete point sources. Although successful in various VLBI studies, it tends to introduce artificial 'clumpiness' in final images, attributed to assumptions about the source structure and effects of sparse $(u,v)$-coverage \citep[e.g.,][]{Lu2023, Savolainen2023}. 

For CLEAN imaging with \difmap{}, we first averaged the data in 10 seconds. We flagged outlier data showing large deviations from neighboring data points. We performed an iterative CLEAN and phase-only self-calibration until the improvement in reduced $\chi^2$ is saturated. Then, an amplitude self-calibration to correct for overall telescope amplitude gain offsets was conducted, followed by an iterative CLEAN and amplitude+phase self-calibration with decreasing solution intervals from 1 hour to 1 minute. At this stage, we averaged the data over all bandwidth (averaging 8 IFs into 1 IF) to increase the SNR, which allows us to perform self-calibration with even shorter solution intervals. The final image was produced when the improvement in reduced $\chi^2$ is saturated in doing an iterative CLEAN and amplitude+phase self-calibration with a solution interval of 10 seconds. For each iteration, we first used a uniform weighting scheme for CLEAN and changed to a natural weighting scheme and performed additional CLEAN to capture the diffuse extended emission. The fluxes of the CLEAN components reconstructed with uniform-weighting are larger than 90\% of the total CLEANed fluxes. We did not use $(u,v)$-tapering to capture more diffuse emission in the further extended jet regions.

Approaches to forward modeling in imaging usually involve representing the image as an array of pixels, requiring a Fourier Transform of this array to assess consistency between the image and data. In RML imaging, the general strategy is to identify the image that minimizes a specified objective function, $\sum \alpha_D \chi^2_D(I) - \sum \beta_R S_R(I)$. Here, $\chi^2_D$ represents a goodness-of-fit function corresponding to the data term $D$, and each $S_R$ is a regularization term corresponding to the regularizer $R$. The hyperparameters $\alpha$ and $\beta$ determine the relative weights between different data terms and regularizers. Using regularizers that emphasize, for example, smoothness, RML imaging can alleviate the clumpiness problem often encountered in CLEAN. Moreover, the output image from RML imaging does not require convolution with a Gaussian beam, as the imaging procedure does not require deconvolution. This highlights that RML methods can achieve a modest degree of "super resolution" (see, e.g., \citealt{Honma2014, Chael2016, Chael2018, Akiyama2017a, Akiyama2017b, EHT2019d, EHT2022c, EHT2024} for more details on RML methods).

For imaging with \ehtim{}, we followed a strategy used in the imaging of the \emph{RadioAstron} space-based VLBI observations of Perseus A \citep{Savolainen2023}. In RML imaging, the MEM regularizer favors pixel-to-pixel similarity to a "prior image." We used the 8.4 GHz CLEAN image NGC 315 observed with the VLBA on January 5, 2020, published in \citetalias{Park2021b}, as the initial prior image. This choice is due to the well-known source structure from previous VLBI observations, aiding imaging convergence. We utilized the lower-frequency image, which lacks detailed information about the fine-scale jet structure due to the limited angular resolution\footnote{The major axis of the synthesized beam is 1.75 mas, which is significantly larger than the transverse size of the jet.}, to mitigate potential bias issues in the imaging procedure. Nevertheless, \ehtim{} could still recover fine-scale jet structures, suggesting the imaging procedure's robustness against potential bias issues.

\begin{figure*}[t]
\centering
\includegraphics[width=0.6\linewidth]{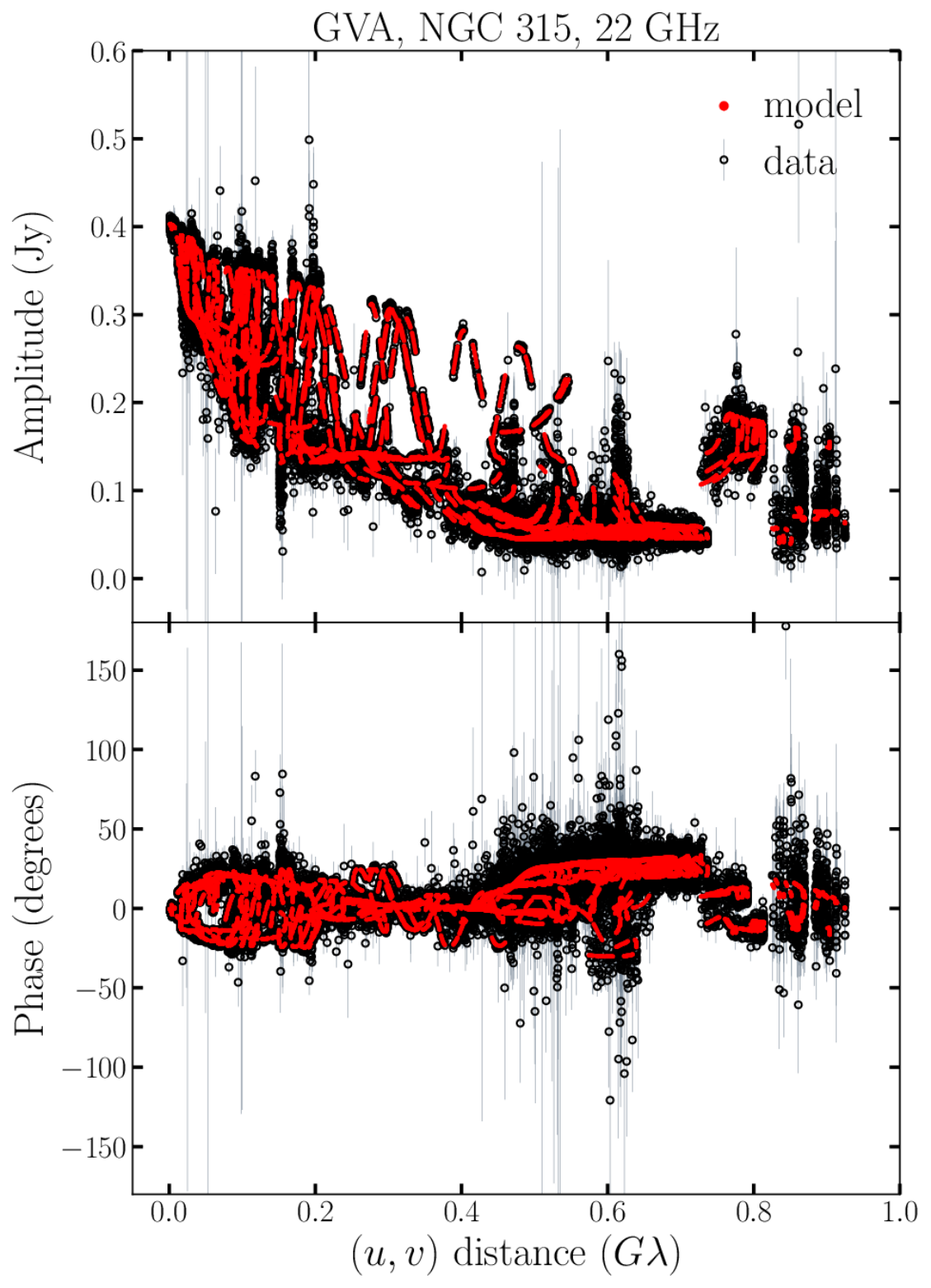}
\caption{Amplitudes (top) and phases (bottom) of the self-calibrated visibilities (black) and RML model (red) are presented as a function of $(u,v)$-distance in units of $10^9$ wavelengths. The data have been averaged over the entire bandwidth and over time in one-minute bins.}
\label{fig:radplot}
\end{figure*}

\begin{figure*}[t]
\centering
\includegraphics[trim = 0mm 0mm 0mm 0mm, width=0.347\linewidth]{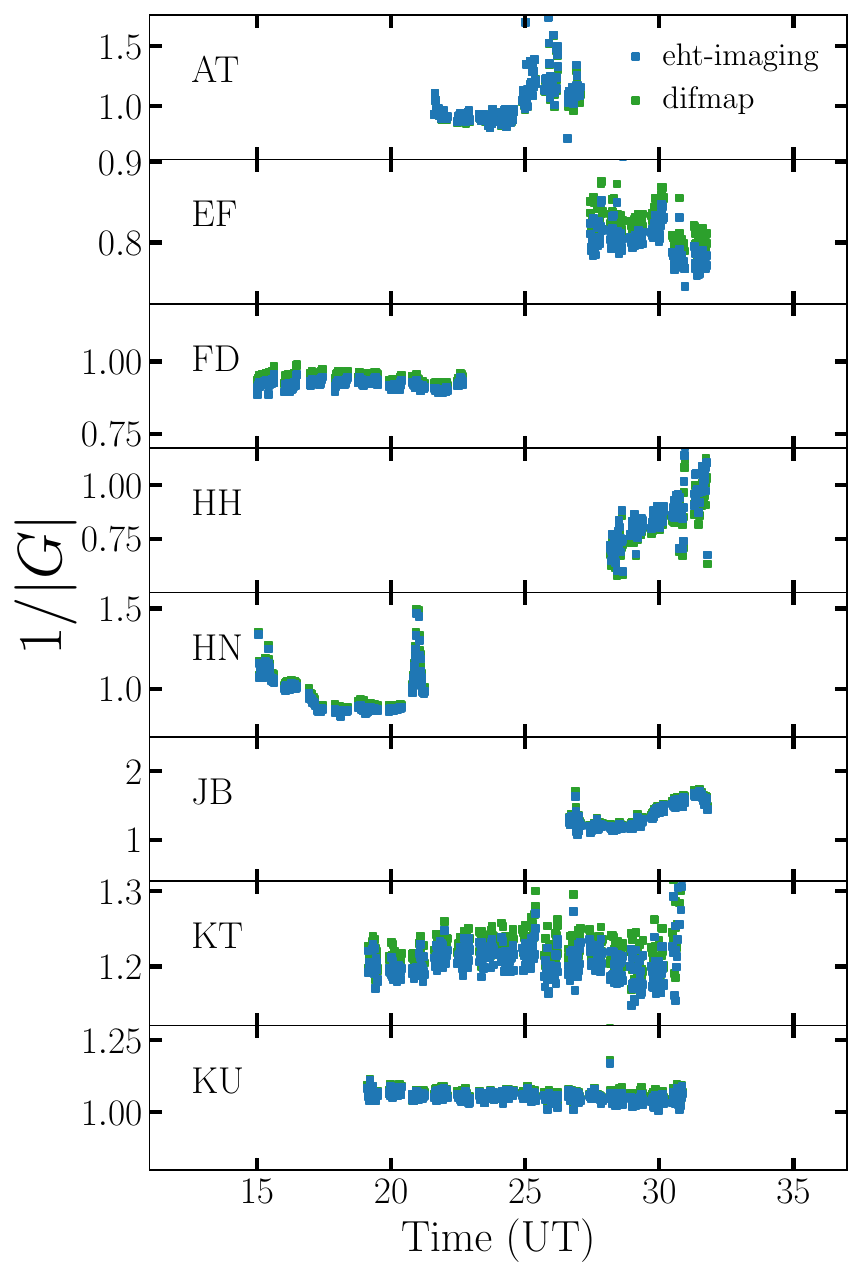}
\includegraphics[trim = 0mm 0mm 0mm 0mm, width=0.32\linewidth]{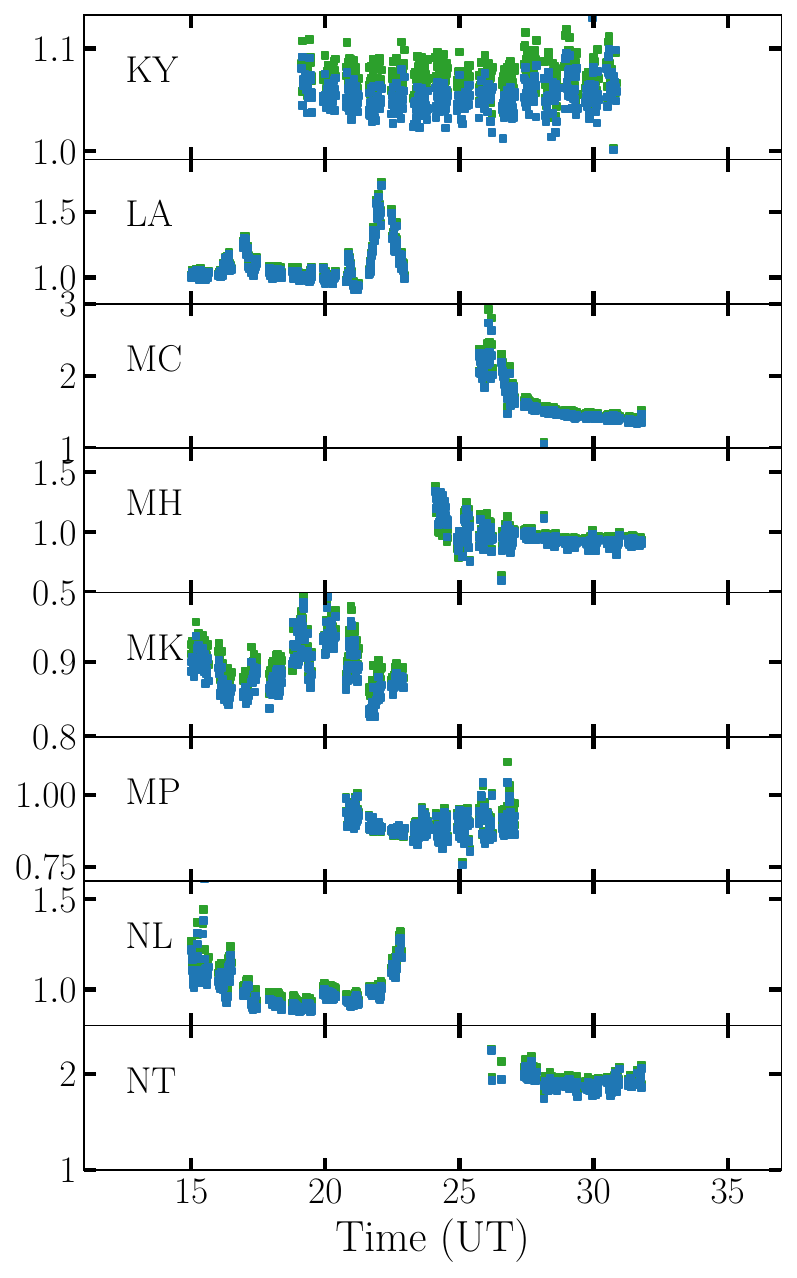}
\includegraphics[trim = 0mm 0mm 0mm 0mm, width=0.313\linewidth]{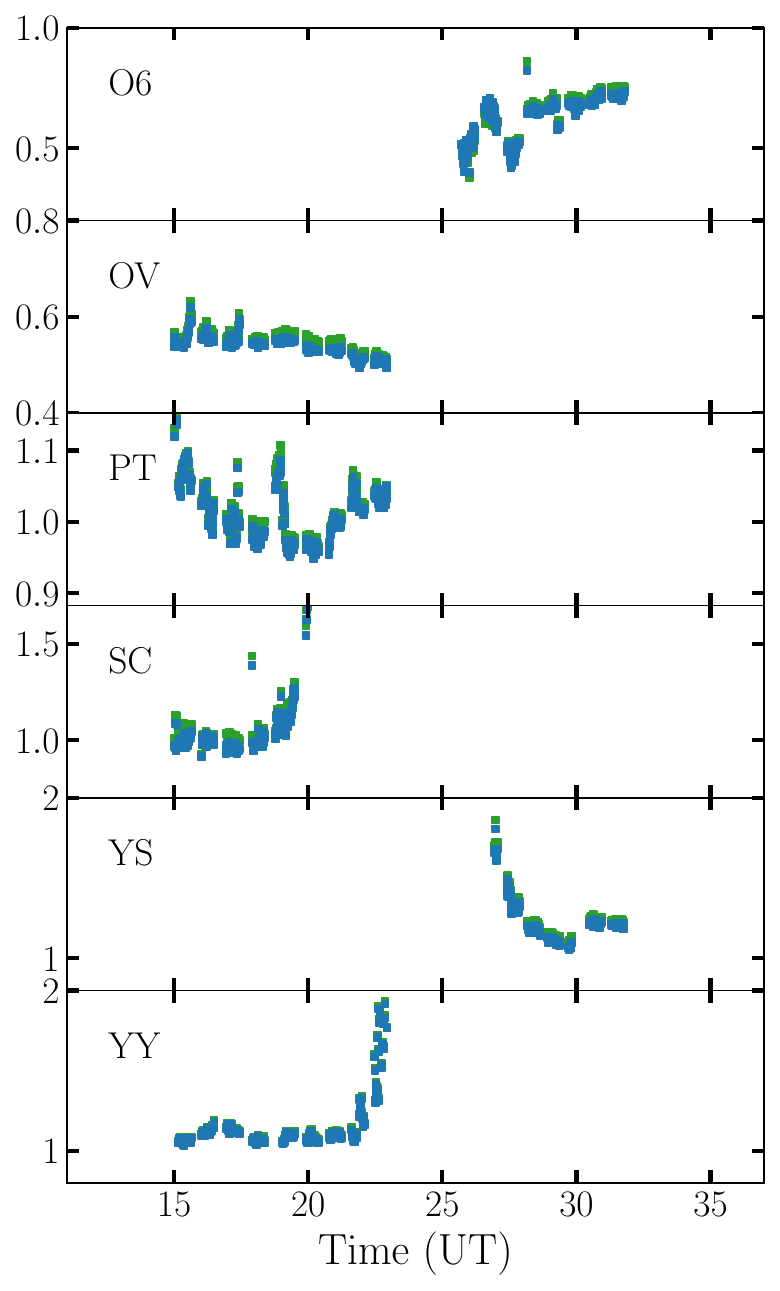}
\caption{Gain correction factors ($1/|G|$) from self-calibration for the representative RML image (blue) and the CLEAN image (green). The factors for the two images are in good agreement with each other. Refer to Table~\ref{tab:antennas} for the stations corresponding to each station code.}
\label{fig:selfcal}
\end{figure*}

We utilized data averaged over frequency and time in 1-minute bins, incorporating a 1.5\% systematic uncertainty\footnote{This means that 1.5\% of the visibility amplitude is added in quadrature to the $1\sigma$ thermal error of the visibility.} \citep{Fuentes2023} to address minor, non-closing systematic errors arising from factors such as polarimetric leakages and uncalibrated antenna gain corruptions within the averaging time bins. In the initial imaging round, data terms comprised closure phases and log closure amplitudes \citep[e.g.,][]{Blackburn2020} exclusively, given that some antennas exhibited significant levels of amplitude gain corruptions, potentially stemming from inaccurate antenna gain curve and system temperature information in the metadata. Following convergence, visibilities underwent self-calibration, and an additional imaging round was conducted, incorporating data terms for closure quantities and visibility amplitudes with a small weight. This imaging/self-calibration process was reiterated up to the sixth round, with a gradual increase in the weight assigned to data terms relative to regularizers, and an increase in the weight for complex visibility in the data terms relative to that for the closure quantities. In the third round, complex visibilities replaced visibility amplitudes in the data terms.

The choice of regularizers and their weights has an impact on the final images. Therefore, we conducted a small parameter survey of $\beta_R$ for four regularizer terms: MEM, TV, TV2, and the $l1$ norm. Additionally, we employed a fixed total flux density regularizer with $\beta_R=10^4$ for the flux density of 0.41 Jy, derived from imaging with \difmap{}, in the initial rounds of imaging. However, $\beta_R=1$ was used in the later stages where the complex visibilities were used for imaging. A detailed description of the parameter survey and a representative set of image reconstructions from various hyperparameter combinations are provided in Appendix~\ref{appendix:survey}. We compared 10 images providing the overall lowest reduced $\chi^2$ for the closure quantities and concluded that \ehtim{} can achieve an effective spatial resolution of 30\% of the diffraction limit (see Appendix~\ref{appendix:survey} for more details). Thus, we blurred all RML images presented in this study with the blurring kernel of the specified size. We selected the image that yields the overall minimum reduced $\chi^2$ for the closure quantities as a representative image. We present the final self-calibrated visibilities and the corresponding model visibilities for the representative RML image in Figure~\ref{fig:radplot}. In Figure~\ref{fig:selfcal}, we present the gain correction factors for the representative RML image and the CLEAN image, which are consistent with each other and behave reasonably.

\begin{figure*}[t]
\centering
\includegraphics[trim = 0mm 0mm 0mm 0mm, width=0.49\linewidth]{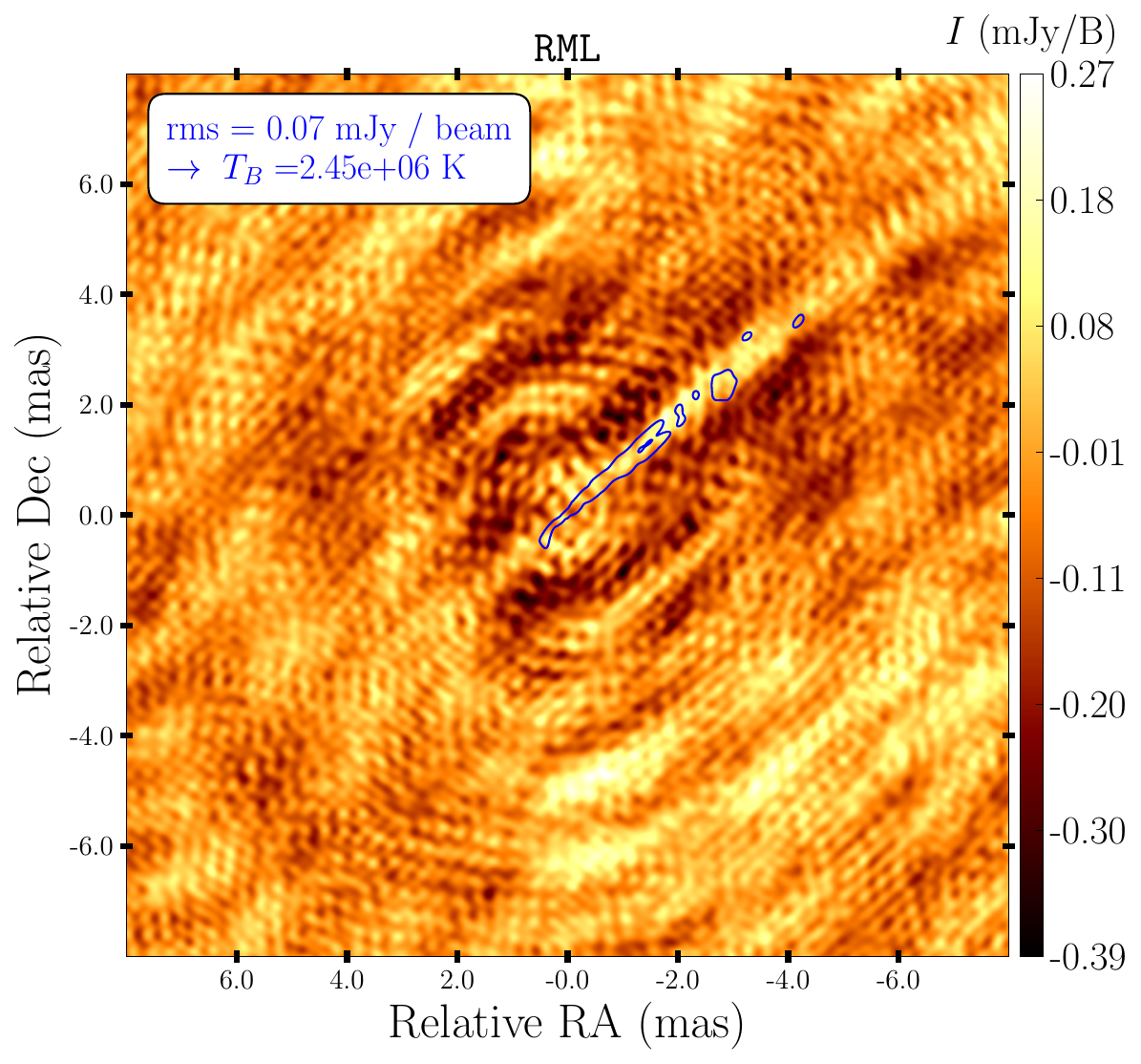}
\includegraphics[trim = 0mm 0mm 0mm 0mm, width=0.49\linewidth]{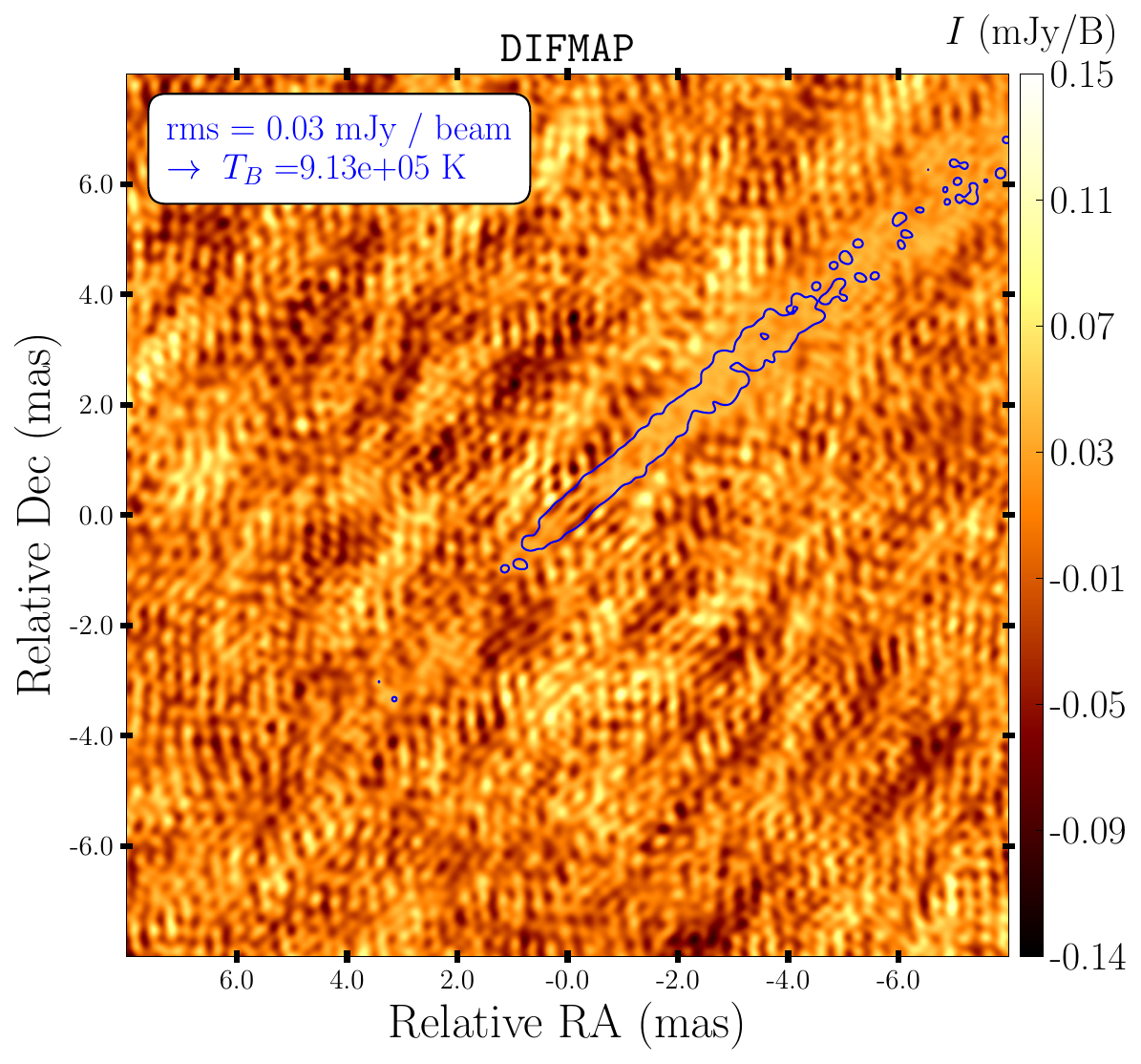}
\caption{Dirty images of the residual visibilities, i.e., the final self-calibrated visibilities subtracted by the model visibilities, for RML (left) and CLEAN (right) models. A natural weighting scheme was used to produce the dirty images. The shape of the observed jet emission in each image (Figure~\ref{fig:fiducial}) is indicated by the white contours. The rms values and the corresponding brightness temperature of the residual images are indicated in each panel.}
\label{fig:dirtymap}
\end{figure*}

To illustrate the robustness of the images, we display the dirty images derived from the residual visibilities—the final self-calibrated visibilities minus the model visibilities—for both the representative RML (left) and CLEAN (right) images in Figure~\ref{fig:dirtymap}. The dirty images suggest two findings: Firstly, the RML residual dirty image exhibits a more pronounced brightness pattern with alternating positive and negative intensities of large amplitudes compared to the CLEAN residual dirty image. This implies that the representative RML image cannot perfectly fit the data, resulting in spurious patterns in the residual dirty images caused by the residual visibilities with non-negligible amplitudes. This result is consistent with the fact that the diffuse extended jet emission at a distance greater than 4 mas from the core is difficult to detect in the RML image (Figure~\ref{fig:fiducial}). However, the RML image can achieve a higher effective resolution than the CLEAN image, enabling the detection of emission with higher brightness temperature, thus achieving a similar image dynamic range to the CLEAN image. Secondly, the rms levels of the RML and CLEAN dirty images are 0.07 and 0.03 mJy per beam, respectively, which translates into brightness temperatures of $2.45\times10^6\ {\rm K}$ and $9.13\times10^5\ {\rm K}$, respectively. The minimum brightness temperature values used for the representative RML and CLEAN images in Figure~\ref{fig:fiducial} are larger than the rms noise levels by factors of 6.5 and 4.4, respectively. Therefore, the jet emission presented in Figure~\ref{fig:fiducial} is significant against the noise levels existing in the residual dirty images.

\section{Imaging Parameter Survey with the \ehtim{} library}
\label{appendix:survey}

In Section~\ref{sec:observation:imaging}, we utilized the RML method from the \ehtim{} library \citep{Chael2016, Chael2018, Chael2023} to image the GVA data. Given that the RML method's output is dependent on the chosen regularizers and their weights, we conducted a small imaging parameter survey to examine the impact of the hyperparameters $\beta_R$. We systematically varied $\beta_R$ from 0 to 100 across four regularizer terms -- MEM, TV, TV2, and the $l1$ norm, following the approach of \citet{Savolainen2023}.

\begin{figure*}[t]
\centering
\includegraphics[width=\linewidth]{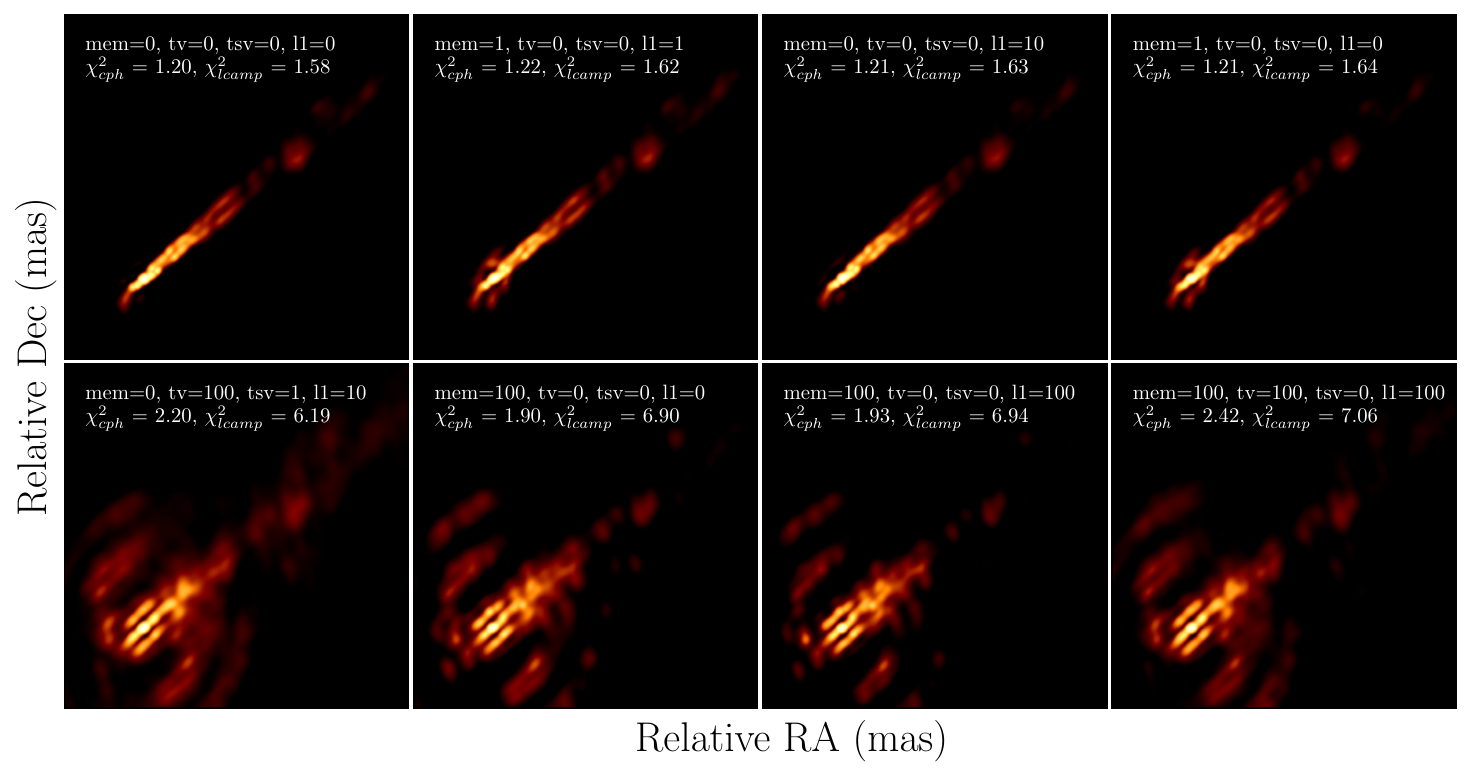}
\caption{Selected RML images from the output of the parameter survey are presented. The weights for each parameter set are indicated in each panel, along with the reduced $\chi^2$ for closure phase ($\chi^2_{cph}$) and log closure amplitude ($\chi^2_{lcamp}$). Images with small reduced $\chi^2$ values closely align with the representative image selected based on the smallest $\chi^2$ values (Figure~\ref{fig:fiducial}). Conversely, images with higher reduced $\chi^2$ values show more pronounced artifacts surrounding the jet emission.}
\label{fig:survey}
\end{figure*}

We present RML images of two groups, one with the smallest $\chi^2$ values and the other with the largest $\chi^2$ values, in Figure~\ref{fig:survey}. The images with the smallest $\chi^2$ values closely resemble the representative image showcased in Figure~\ref{fig:fiducial}, distinctly exhibiting a limb-brightened jet morphology. Conversely, the images with the largest $\chi^2$ values display more pronounced brightness patterns surrounding the jet emission. These patterns are considered imaging artifacts, as they fail to adequately explain the observed closure quantities.

It is known that RML imaging methods can achieve "super resolution," exceeding the nominal diffraction limit ($\theta\sim\lambda/|u|_{\rm max}$ with $|u|_{\rm max}$ being the maximum baseline length of the VLBI data used). The extent of super resolution varies based on the data and source model, but previous studies have shown that these methods can attain an effective spatial resolution of $\approx20-50$\% of the diffraction limit \citep{Chael2016, Akiyama2017b}. In this study, we determined the effective spatial resolution as follows: we selected 10 images with the lowest overall closure chi-squares, assuming these images most accurately represent the true source model (Figure~\ref{fig:survey}). We then blurred these images with a circular Gaussian blurring kernel with a FWHM ranging from 0.1 to 1.0 of the diffraction limit. For each kernel size, we computed the normalized cross-correlation:
\begin{equation}
\rho_{\text{NX}}(X,Y) = \frac{1}{N}\sum_i \frac{(X_i - \langle X \rangle)(Y_i - \langle Y \rangle)}{\sigma_X \sigma_Y},
\end{equation}
where the sum is over all $N$ pixels in the two images, $\langle{X}\rangle$ and $\langle{Y}\rangle$ are the mean pixel values in the images, and $\sigma_X$ and $\sigma_Y$ are the standard deviations of the pixel values in each image. This calculation was performed using \ehtim{}, and we computed the mean $\rho_{\rm NX}(X,Y)$ for all possible combinations of the selected images for each blurring kernel size.

We observed that $\rho_{\rm NX}(X,Y)$ increases with larger blurring kernel sizes, as expected. However, the rate of increase significantly flattens at a blurring kernel size of 0.3 of the diffraction limit (the mean $\rho_{\rm NX}(X,Y)$ reaches approximately 0.98 at this kernel size). Based on this finding, we conclude that \ehtim{} can achieve an effective spatial resolution of 30\% of the diffraction limit, and accordingly, we blurred all RML images presented in this study with the blurring kernel of this specified size.

\section{Image Dependence on VLBI Arrays}
\label{appendix:jackknife}

\begin{figure*}[t]
\centering
\includegraphics[trim = 0mm 5mm 0mm 0mm, width=\linewidth]{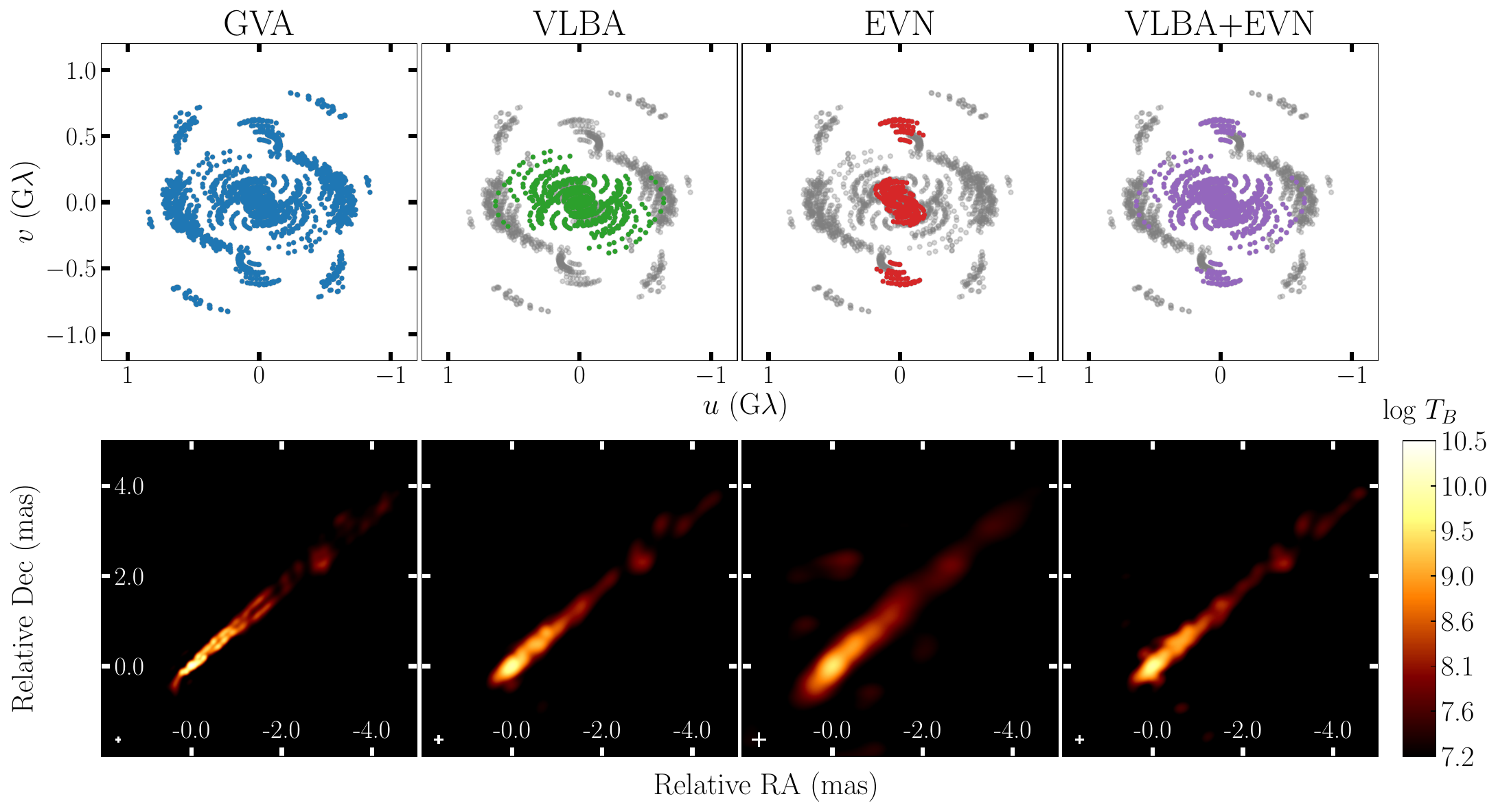}
\caption{\emph{Top:} $(u,v)$-coverages of the original GVA data, the VLBA and phased VLA, the EVN, and the VLBA+EVN, from left to right. We overplot the $(u,v)$-coverage of the GVA data in light grey for easier comparison. \emph{Bottom:} Reconstructed images from each array using \ehtim{}. In the lower left corner of each image, we indicate the size of the blurring kernel for each image, determined by computing $\rho_{\rm NX}(X,Y)$ for all possible combinations of the selected 10 images with the lowest overall $\chi^2$ values for each blurring kernel size, as done in Appendix~\ref{appendix:survey}. The limb-brightening of the NGC 315 jet is successfully reconstructed only with the full GVA array.}
\label{fig:jackknife}
\end{figure*}

In this Appendix, we present the results of imaging data from parts of the GVA only: the VLBA (including the phased VLA), the EVN, and the VLBA+EVN. Note that the imaging simulations did not include stations that did not participate in the GP060 observation, such as the Russian Kvazar and Chinese stations in the EVN. The imaging reconstruction was performed using \ehtim{}, following the same methodology as described in Section~\ref{sec:observation:imaging}. One difference is that we conducted a parameter survey for each dataset for 30 parameter sets that provided the overall lowest closure $\chi^2$ for the GVA data to reduce computation time. Figure~\ref{fig:jackknife} displays the $(u,v)$-coverage of each dataset and the corresponding image reconstructions that provided the lowest overall $\chi^2$ values.

The reconstructed images from each VLBI array show two-sided jets, consistent with those in previous studies (e.g., \citetalias{Park2021b}, \citealt{Boccardi2021, Ricci2022}). However, there is no clear evidence of limb-brightening. This outcome suggests that the robust detection of limb-brightening is facilitated by the extensive $(u,v)$-coverage provided by the GVA, particularly with the inclusion of the Australian LBA and the KVN.

\section{Discussion on Jet Acceleration and Collimation Zone Discrepancy in Previous NGC 315 Studies}
\label{appendix:boccardi}

A discrepancy exists in the reported size of the jet acceleration and collimation zone for NGC 315 between previous studies (\citetalias{Park2021b}; \citealt{Boccardi2021, Ricci2022}). \citetalias{Park2021b} reported that the jets in NGC 315 are accelerated and collimated at distances down to approximately $10^5\ R_g$, whereas other studies \citep{Boccardi2021, Ricci2022} suggest that the jet acceleration and collimation complete at approximately $10^4\ R_g$. The cause of this discrepancy remains unclear, and in this appendix, we briefly summarize potential reasons for it.

\begin{figure}[t]
\centering
\includegraphics[width=0.6\linewidth]{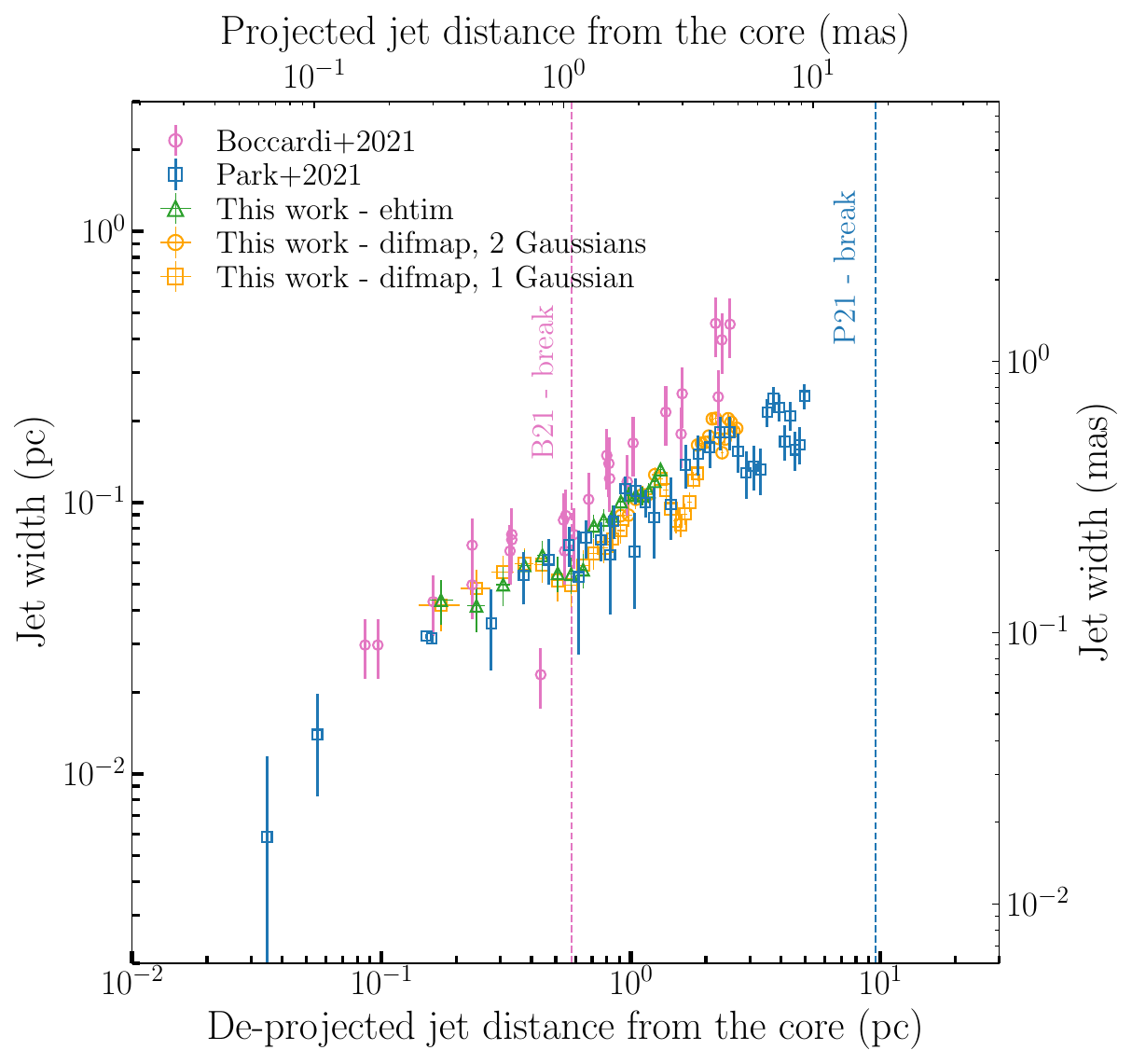}
\caption{Jet width as a function of de-projected distance from the core is shown, derived from the RML image (green triangles) and the CLEAN image (yellow circles and squares), as well as from the previous studies \citetalias{Park2021b} (22 and 43 GHz; blue squares) and \cite{Boccardi2021} (22, 43, and 86 GHz; pink circles). The length units are in pc for the bottom and left axes and in mas for the top and right axes. We used the same conversion factor of 1 mas to 0.331 pc and a jet viewing angle of $38^\circ$ for de-projection, as in \cite{Boccardi2021}. The locations of the jet collimation breaks observed in those studies are indicated by the vertical dashed lines. Note that the distance from the core is specified at each frequency, as no core-shift effect was corrected.}
\label{fig:boccardi}
\end{figure}

\begin{itemize}
    \item \textbf{Data.} \citetalias{Park2021b} utilized simultaneous multifrequency VLBA data spanning 1.5 to 22.2 GHz, along with one archival HSA dataset at 43.2 GHz and archival VLA data at 1.4 and 4.9 GHz. In contrast, \cite{Boccardi2021, Ricci2022} employed various archival VLBA/HSA datasets, conducted their observations at 22, 43, and 86 GHz, and utilized archival VLA 1.4 and 4.9 GHz data. \cite{Kovalev2020} suggested that VLBI images with limited sensitivity could affect the jet width measurement, as the jet emission may not cover the full jet width due to the sensitivity limitation.
    \item \textbf{Analysis methods.} The most significant difference in analysis methods lies in how to derive the jet width. \citetalias{Park2021b} obtained the jet width in the image domain by fitting a single Gaussian function to the transversely sliced intensity profile at each distance bin, while \citet{Boccardi2021} used the \texttt{modelfit} subroutine implemented in \difmap{} to fit a set of circular Gaussian components directly to the visibilities. The former approach\footnote{We note that this method is commonly used in studies of jet collimation profiles \citep[e.g.,][]{Tseng2016, Akiyama2017a, Pushkarev2017, Nakahara2018, Nakahara2019, Kovalev2020, Baczko2022, Okino2022, Yan2023}.} is likely more influenced by the synthesized beam of the data, as it derives the jet width from the image produced by restoring the CLEAN models with the beam. Conversely, the latter method is influenced by the simple assumption that the jet structure is described by components with a pre-defined brightness distribution (2D circular Gaussian), which might be less accurate for the true, potentially more complex, jet structure.
    \item \textbf{Core-shift measurements.} Although the same method was used to derive the core-shift in \citetalias{Park2021b} and \citet{Boccardi2021}, based on 2D cross-correlation of optically thin jet emission at different frequencies \citep{CG2008}, the reported core-shift amounts in the studies differ significantly. This discrepancy may lead to substantial differences in the derived jet acceleration profile (\citetalias{Park2021b} and \citealt{Ricci2022}). The jet speeds were determined based on the intensity ratio between the jet and counterjet, which can be significantly affected by the reference point assumed for the origin of the jet and counterjet, determined using core-shift measurements. Note that the core-shift itself can significantly vary over time \citep[e.g.,][]{Lisakov2017, Plavin2019}, which could serve as another source of discrepancy between the studies.
\end{itemize}

Although a dedicated study is required for a comprehensive understanding of the origins of the discrepancy, our GVA observation can resolve the jet transversely, enabling more robust measurements of the jet width. This is valuable for investigating the origin of the discrepancy in the jet collimation profile. In Figure~\ref{fig:boccardi}, we compare the jet width measured in the present study using the GVA data with those reported in \citetalias{Park2021b} and \cite{Boccardi2021}. We excluded any core-shift effects from the data in the figure, given the differing reported core-shift amounts in those studies. This allows for an apples-to-apples comparison. Jet widths were compared at high frequencies (22, 43, and 86 GHz for \citealt{Boccardi2021} and 22 and 43 GHz for \citetalias{Park2021b}). The core-shift effect among these frequencies is small, allowing us to reasonably ignore the difference in the jet collimation profile at those frequencies caused by the core-shift effect. We used the same jet viewing angle ($38^\circ$; \citealt{Canvin2005}) and length unit (pc) as in \cite{Boccardi2021} to ensure reproducibility of the jet collimation profile presented in that paper.

We found that the jet widths measured in the present study are consistent with both \citetalias{Park2021b} and \citet{Boccardi2021} at sub-parsec scales. However, the discrepancy becomes more pronounced at larger distances. The GVA data did not show a jet shape transition from a parabolic to a conical shape at the subparsec scale, as reported in \citet{Boccardi2021}. The origin of the discrepancy between our GVA results and those reported in \citet{Boccardi2021} is still puzzling. One plausible explanation is due to the temporal variability of the jet structure, since the data from \citetalias{Park2021b} and this study are post-2020, whereas those in \citet{Boccardi2021} are predominantly pre-2010. We plan to investigate the origin of the discrepancy in a forthcoming paper (Ricci, L. et al. in prep.).

\end{appendix}

\bibliography{main}{}
\bibliographystyle{aasjournal}

\end{document}